\documentclass[aps,prb,twocolumn,groupedaddress,showpacs,superscriptaddress,amssymb,amsmath]{revtex4-1}

\usepackage{graphicx}
\usepackage{color}
\usepackage{amsmath}
\usepackage{comment}
\usepackage{placeins}
\newcommand{\be}{\begin{equation}}
\newcommand{\ee}{\end{equation}}
\newcommand{\bea}{\begin{eqnarray}}
\newcommand{\eea}{\end{eqnarray}}
\usepackage{dcolumn}
\usepackage{bm}
\usepackage{hyperref}
\usepackage{epsf}
\usepackage{subfigure}
\usepackage{epstopdf}%
\usepackage{amsfonts}

\begin{document}
%

\title{Full counting statistics of vibrationally-assisted electronic conduction:
Transport and fluctuations of the thermoelectric efficiency}

\author{Bijay Kumar Agarwalla}
\affiliation{Chemical Physics Theory Group, Department of Chemistry, and
Centre for Quantum Information and Quantum Control,
University of Toronto, 80 Saint George St., Toronto, Ontario, Canada M5S 3H6}

\author{Jian-Hua Jiang}
\affiliation{Department of Physics, Soochow University, 1 Shizi Street, Suzhou 215006, China}

\author{Dvira Segal}
\affiliation{Chemical Physics Theory Group, Department of Chemistry, and
Centre for Quantum Information and Quantum Control,
University of Toronto, 80 Saint George St., Toronto, Ontario, Canada M5S 3H6}

\date{\today}


\begin{abstract}
We study the statistical properties of charge and energy transport in
electron conducting junctions with electron-phonon
interactions, specifically, the thermoelectric efficiency and its fluctuations.
The system comprises donor and acceptor electronic states, representing a two-site molecule
or a double quantum dot system.
Electron transfer between metals through the two molecular sites is coupled to a particular vibrational mode
which is taken to be either harmonic or anharmonic- a truncated (two-state) spectrum. 
Considering these models 
we derive the cumulant generating function in steady state
for charge and energy transfer, correct  to second-order in the
electron-phonon interaction, but exact to all orders in the metal-molecule coupling strength.
This is achieved by
using the non-equilibrium Green's function approach (harmonic mode) and
 a kinetic quantum master equation method (anharmonic mode).
From the cumulant generating function
we calculate the charge current and its noise and
the large deviation function for the thermoelectric efficiency.
%
We demonstrate that at large bias the charge current, differential conductance, and the current noise
 can identify energetic and structural properties of the junction.
We further examine the operation of the junction as a thermoelectric engine and show that
while the
macroscopic thermoelectric efficiency is indifferent to the nature of the mode (harmonic or anharmonic),
efficiency fluctuations do reflect this property.

\end{abstract}

\maketitle

\section{Introduction}
\label{Intro}

Single-molecule junctions offer a versatile playground for probing basic questions in condensed phases physics:
How do quantum effects and many-body interactions (electron-electron, electron-phonon)
control charge and energy transport processes, thus the operation of nano-scale atomic and molecular devices
\cite{forty,Latha13}?
How do we accurately and efficiently simulate quantum transport phenomena involving different particles
(and quasi-particles), electrons, phonons, spins, polarons?
Recent progress in experimental techniques has made it possible to perform sensitive measurements at the molecular
scale, in the linear and non-linear transport regimes, to
observe signatures of many body effects.
Kondo physics, the hallmark of strongly correlated electrons, was observed in different molecules,
see e.g. Ref. \cite{KondoNatelson}.
Coupled electron-vibration processes were probed in single-molecule junctions applying
inelastic electron tunneling spectroscopy \cite{NatelsonIETS,Tal13,Tal14} and Raman spectroscopy tools
\cite{Nat-Raman,Selzer,Natelson14},
displaying frequency shifts and mode heating 
in response to electron conduction.
Noise characteristics of the charge current can further
expose the nature of the vibrational modes contributing to electron dynamics \cite{Tal08,Ruitenbeek,Tal13}.

Theoretical and computational methodologies dedicated
to the effects of electron-phonon interactions on transport in nano-conductors
were reviewed in Refs. \cite{Galp07,luAIP}.
The celebrated Anderson-Holstein model, with a
single electronic orbital coupled to a local phonon mode,
exposes an intricate interplay between the electronic and nuclear degrees of freedom.
The model has been extensively studied to reveal the behavior of the current and its fluctuations in different regimes
of electron-phonon coupling,
see for example  \cite{galperinIN06, galperinIN07, Schmidt,Ora09, Levy1, Levy2,Levy3,Levy4,Levy5,Gogolin,thorwart,Thoss1,Simine14,Utsumi1,Ness,Joe2,Joe3}.
An extension of the  Anderson-Holstein model with a secondary phonon bath was examined in many studies,
see e.g. the exploration of thermoelectric transport in a three-terminal junction
\cite{Ora10} and the analysis of transient effects \cite{Rabani14}.

\begin{figure*}
\includegraphics[width=12cm]{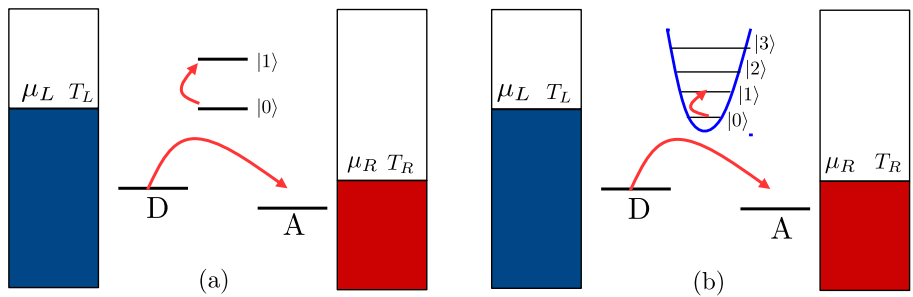}
\caption{(color online)
Scheme of a donor (D) - acceptor (A) molecular junction.
(a) In the DA-AH model electron transfer is coupled to a highly anharmonic
vibrational mode which consists of two levels.
(b) In the DA-HO model the vibrational mode is assumed harmonic.
The arrows exemplify an inelastic coupled transport process,
with an electron hopping from left to right, assisted by
an excitation of the mode.
In the analysis of thermoelectric devices we set $T_R>T_L$,  $\mu_L> \mu_R$.}
\label{figS}
\end{figure*}


Complementing the Anderson-Holstein model,
the donor-acceptor (DA) prototype junction of Fig. \ref{figS} allows the exploration of a broad range of
problems. The model comprises two electronic states, referred to as ``donor" (D)
and ''acceptor" (A), following the
chemistry literature on electron transfer reactions.
Electron transfer between the D and A sites is
assisted by a particular vibrational mode, isolated, or coupled to a secondary phonon bath.
This model, suggested to describe a molecular electronic rectifier \cite{Ratner-Aviram},
was recently revisited and analyzed using a variety of tools, for example, the
Fermi golden rule approach \cite{Brandbyge,Ora13}, quantum master equations (QME)
\cite{Felix,SegalSF}, the non-equilibrium Green's function (NEGF) technique
\cite{Thoss2,Galperin-Park,Galperin-pseudo} and from influence functional path integral simulations
\cite{SegalINFPI}. 
In molecules, the DA model represents
charge transfer between weakly connected chemical groups, facilitated by a  vibrational mode  \cite{Brandbyge}.
In the context of nanoelectromechanical systems \cite{Blencowe}
the two electronic states can be realized by
gate-controlled quantum dots which are coupled by a mechanical system \cite{Dima},
a suspended tunnel-junction such as a carbon nanotube.
Electrons transferred between the quantum dots can e.g. excite transverse acoustic modes of the suspended
tube  \cite{ilani14,Felix}.
%
The single bosonic mode can also represent a cavity mode assisting electron tunneling between quantum dots,
and other hybrid models \cite{Nori, Vahala}.

The DA model in Fig. \ref{figS} offers a rich setting for
investigating the role of inelastic-vibrationally assisted electron scattering
in far-from-equilibrium (nonlinear) situations. 
In contrast to coherent conduction,
inelastic electron transport can realize nontrivial effects beyond linear response,
such as charge and thermal rectification and cross-rectification effects. 
This was e.g. demonstrated in Ref. \cite{Jiang} within a three-terminal DA junction, by
replacing the single vibration by a phonon bath.

Traditionally, quantities of interest in transport experiments and calculations
were averaged values for population and currents.
However, it should be recognized that nanoscale junctions
suffer from strong random fluctuations due to their surrounding environments.
It is therefore desirable to develop a probabilistic theory for measurable quantities.
Indeed, in small systems the second law of thermodynamics
should be replaced by a universal symmetry, the fluctuation theorem \cite{Jarzynski, Esp5-review, Hanggi-review},
\be
\frac{P_t(S)}{P_t(-S)} = e^S,
\ee
where $P_t(S)$ ($P_t(-S))$ is the probability distribution for observing a positive (negative)
entropy production during the time interval $t$.

Fluctuations in heat provided to a nanoscale engine
and work performed naturally translate to stochastic efficiencies.
Universal characteristics of the corresponding probability distribution function $P_t(\eta)$,
for time-reversal symmetric engines, were explored in Refs. \cite{Esp2,Esp3,Esp4},
from the principles of classical stochastic thermodynamics.
It can be proved that the large deviation function (LDF) for efficiency,
defined as $\tilde J(\eta)= - \lim_{t\to \infty} t^{-1} \ln [P_t(\eta)]$,
attains a global minimum which coincides with the macroscopic (average) efficiency,
and a global maximum which corresponds to the least probable efficiency,
coinciding with the Carnot efficiency \cite{Esp2,Esp3,Esp4}.
These universal features are a direct consequence of the fluctuation theorem.
Extensions of this analysis to explore efficiency statistics for systems with broken time-reversal symmetry
were given in Ref. \cite{Hua-Bijay}.
Fluctuations of the finite-time efficiency were experimentally demonstrated in Ref. \cite{expt-fluc}.
Beyond classical thermodynamics, in a recent study the concept of the
``stochastic efficiency" was examined within a quantum coherent model
of a thermoelectric junction, by employing the non-equilibrium Green's function technique \cite{Esp1}.

In this paper, we provide a complete analysis of charge and energy transport behavior in the
donor-acceptor junction by
taking a full counting statistics (FCS) approach.
We consider two variants of the model as depicted in Fig. \ref{figS}:
Electron transfer may couple to a harmonic vibrational mode (DA-HO model),
or to an anharmonic impurity (DA-AH model).
The latter case is represented by a two-state system, a truncated vibrational manifold.

We rigorously derive the cumulant generating functions for the models of Fig. \ref{figS}
under the assumption of weak electron-vibration coupling,
handing over the complete information over the models' steady state transport behavior.
From the cumulant generating functions we explore
transport in the junctions far from equilibrium, 
specifically, we aim in identifying transport quantities which are sensitive to the nature of the vibrational mode.
Finally, we investigate thermoelectric efficiency fluctuations in the DA junction beyond linear response.
Interestingly, our analysis in this paper exemplifies that one can reconcile
two central yet disparate techniques, QME and NEGF, to obtain consistent results, within the same order in perturbation theory.

The paper in organized as follows.
We introduce the DA junction in Section \ref{Model} and
perform a FCS analysis in Sec. \ref{FCS}.
The cumulant generating function (CGF) of the DA-AH model is derived in Sec. \ref{FCS-AH} by
employing the quantum master equation approach. The derivation of the CGF
for the DA-HO case is detailed in Sec. \ref{FCS-HO},
using the non-equilibrium Green's function technique.
Two applications are described in Sec. \ref{Applications}:
We simulate the junction's current-voltage characteristics in Sec. \ref{App-IV}, and examine
the statistics of the thermoelectric efficiency in Sec. \ref{App-eta},
further comparing numerical results far from equilibrium with the linear response (Gaussian) limit.
Our findings are summarized in Sec. \ref{Summ}.
For simplicity, we set $e=\hbar=k_B=1$ throughout derivations.


\section{Model}
\label{Model}

The DA junction includes a two-site structure, representing a donor-acceptor molecule
(or equivalently, a double-quantum-dot system),
placed in between two metal leads.
The total Hamiltonian is given by
\be
H_T= H_M+ H_L+ H_R + H_C + H_{vib} + H_{I}.
\label{Full-H}
\ee
The molecular Hamiltonian $H_M$ includes the donor and acceptor sites,
\be
H_M= {\epsilon}_d c_d^{\dagger} c_d + {\epsilon}_a c_a^{\dagger} c_a,
\ee
with $c_{d/a} (c_{d/a}^{\dagger})$ as a fermionic annihilation (creation) operator at the donor or
acceptor sites with energies ${\epsilon}_{d/a}$.
The second and third terms in
Eq.~(\ref{Full-H}) represent the left ($H_L$) and right ($H_R$) metal leads,
modeled by collections of non-interacting electrons,
\be
H_{L}= \sum_{l\in L} {\epsilon}_l c_l^{\dagger} c_l, \quad H_{R}= \sum_{r\in R} {\epsilon}_r c_r^{\dagger} c_r,
\ee
with the fermionic annihilation (creation) operators $c_{j}$ ($c_j^{\dagger})$.
The tunneling energies of electrons from the donor (acceptor) site to the left (right) lead  $v_{l}$ ($v_r$)
are included in $H_C$, and are assumed to be real valued,
\be
H_{C}= \sum_{l \in L} v_{l} (c_l^{\dagger} c_d + c_d^{\dagger} c_l) + \sum_{r \in R} v_{r} (c_r^{\dagger} c_a + c_{a}^{\dagger} c_r).
\label{eq:HC}
\ee
$H_{vib}$ and $H_{I}$ represent the Hamiltonians of the molecular vibrational mode and
its coupling with the D and A sites (strength $g$). 
Assuming an harmonic local mode we write
\bea
H_{vib}&=& \omega_{0} b_{0}^{\dagger} b_0,
\nonumber \\
H_{I} &=& g [c_{d}^{\dagger} c_{a} + c_{a}^{\dagger} c_d ]  (b_0^{\dagger} + b_0),
\label{eq:HI}
\eea
with $b_0(b_0^{\dagger})$ a bosonic annihilation (creation) operator for a vibrational mode of frequency
$\omega_0$.  
The interaction  $H_I$ is sometimes refereed to as an ``off-diagonal'' 
since electron exchange between the two sites is allowed only via the excitation and/or
relaxation of the mode, see Fig. \ref{figS}.
Note that we do not include here a direct tunneling (elastic) term between sites D and A.
This simplification allows us to reach closed analytic results for the CGF.
The contributing of elastic tunneling processes can be included in an additive manner
\cite{Jiang}, a reasonable approximation at weak coupling \cite{SegalINFPI}.

We diagonalize the electronic Hamiltonian,
$H_{\rm el}\equiv H_M+ H_L+ H_R + H_C$,
and write it down in terms of a new set of fermionic operators $(a_{l/r},a^{\dagger}_{l/r})$,
\be
H_{\rm el} = \sum_{l} \epsilon_l a_l^{\dagger} a_l + \sum_{r} \epsilon_r a_r^{\dagger} a_r.
\label{eq:Hel}
\ee
These operators are related to the original set by \cite{SegalSF}
\bea
c_d &=& \sum_{l} \gamma_l a_l, \quad c_a = \sum_{r} \gamma_r a_r, \nonumber \\
c_l &=& \sum_{l'} \eta_{l l'} a_{l'}, \quad c_r =\sum_{r'} \eta_{r r'} a_{r'},
\label{eq:newoperators}
\eea
with the dimensionless coefficients
\be
\gamma_l = \frac{v_l}{\epsilon_l-\epsilon_d - \sum_{l'} \frac{v_{l'}^2}{\epsilon_l-\epsilon_{l'} + i \delta}},
\quad \eta_{ll'} = \delta_{ll'}- \frac{v_l \gamma_{l'}}{\epsilon_l-\epsilon_{l'} + i \delta}.
\label{eq:trans}
\ee
%
Here $\delta$ is a positive infinitesimal number introduced to ensure causality.
Analogous expressions hold for the $r$ set.
The expectation values of number operators obey, e.g., at the $L$ end,
$ \langle a_{l}^{\dagger} a_{l'} \rangle = \delta_{l l'} f_L(\epsilon_{l})$
with $f_{L}(\epsilon_{l})= \big\{\exp[\beta_{L} (\epsilon_{l}-\mu_{L})] +1\big\}^{-1}$
the Fermi distribution for the lead with chemical potential $\mu_{L}$ and an inverse temperature $\beta_{L}$.
Using the new operators,
we write down the total Hamiltonian for the DA-HO junction as
\bea
H_{DA-HO}&=& H_{\rm el} + \omega_0 b_0^{\dagger} b_0
\nonumber\\
&+& g \sum_{l\in L,r\in R} \big[\gamma_l^* \gamma_r a_l^{\dagger} a_r + h.c.\big](b_0^{\dagger} + b_0).
\label{eq:DA-HO}
\eea
The last term in the Hamiltonian describes electron-hole pair generation
assisted by the interaction with the vibrational mode.

In a simpler version of this model we replace the infinite level spectrum of
the harmonic oscillator by a truncated two-level system, to mimic a highly anharmonic vibrational mode.
The Hamiltonian of this DA-AH model can be conveniently written in terms of the Pauli matrices as
\be
H_{DA-AH}= H_{\rm el} + \frac{\omega_0}{2} \sigma_z + g \sum_{l\in L,r\in R} \big[\gamma_l^* \gamma_r a_l^{\dagger} a_r + h.c.\big]\sigma_x.
\label{eq:DA-AH}
\ee
%

The DA-HO and DA-AH models described above are simple enough to allow us to derive expressions for the corresponding CGFs.
Meanwhile, we  (i) reconcile the NEGF approach with QME, 
to assist in method development in the area of quantum transport, and (ii)
provide analytic results for transport characteristics, to bring intuitive guidelines for functionality.

We substantiate our modeling by describing connections to ab-initio studies of molecular conduction \cite{SCBA}.  
The donor and acceptor sites could represent spatially separated electronic states in 
a molecule or atoms in an atomic wire, with
the electron-phonon interaction of Eq. (\ref{eq:HI}) describing a bond-length stretching mode.
The rigid motion of a molecule/atomic chain between the metal leads
can be modeled by an additional electron-phonon interaction Hamiltonian 
$H_I^{(1)}=g_1(b_1^{\dagger}+b_1)(c_d^{\dagger}c_d+c_a^{\dagger}c_a)$. This mode is not
included in the present study.
As well, we ignore direct tunneling between electronic sites in the form 
$H_{tunn}=v_{d,a}\left(c_d^{\dagger}c_a+c_a^{\dagger}c_d\right)$.
Detailed ab-initio studies of electron-phonon inelastic effects in molecular junctions prepare direct tunneling elements,
frequencies of active modes, and their 
electron-phonon matrix elements, 
see e.g. Refs. \cite{Fred04,SCBA,JauhoR,Guo,Fred07}.
Certain type of molecular junctions 
could be well represented by our model, when electron transport due to $H_{I}$ dominates over both elastic tunneling 
$H_{tunn}$ and the diagonal electron-phonon coupling  $H_{I}^{(1)}$. This is the case
e.g. in Ref. \cite{Markussen}, considering
charge transfer through a biphenyl molecule with the torsion motion assisting electron hopping between the (almost orthogonal) benzene rings.
Particularly,  when $\epsilon_d\neq \epsilon_a$, calculations of transport in double quantum dot systems show that
the inelastic component of the current can dominate the elastic term \cite{Jiang}.

We also justify our model in the language of
molecular orbitals, electronic eigenstates of the molecule. 
Consider two orbitals, each coupled to both metal leads- but in an asymmetric manner:
one orbital couples strongly to 
the left lead but weakly to the right, the other molecular orbital is strongly coupled to the right lead but weakly to the left side.
In the absence of electron-phonon interaction this molecule supports very small currents. 
It will however turn into a good conductor
at high enough temperatures when phonons contributing to  (\ref{eq:HI}) are active,
supporting inelastic current. For an extended discussion, see Ref. \cite{thoss15}.

Turning to the the AH model, the two-state impurity
describe deviations from the harmonic picture. Besides electron-phonon coupled situations, the model
could represent electron transport junctions interacting with a local spin impurity, see e.g. Refs. \cite{mag1,mag2},
demonstrating electronic read-out of nuclear spins.

We finally comment that in the non-crossing approximation,
when elastic and inelastic tunnelling events do not interfere \cite{Mahan,Galperin-Park},
the current (or more generally, the cumulant generating function) can be written as a sum of elastic and inelastic terms.
While here we treat the inelastic component only,
the CGF for elastic transport is well known \cite{Levitov},
bringing in the standard Landauer formula for currents. 

In the next Section we develop a counting statistics approach for charge and energy transfer processes.
We then analyze first the (simpler) DA-AH model using a QME approach, followed by the investigation of the DA-HO junction by
utilizing the NEGF technique.

\section{Counting statistics for charge and energy currents}
\label{FCS}

The cumulant generating function contains information over the statistics of transferred particles and energy flowing across the system,
potentially far from equilibrium.
Here we are interested in the CGF for coupled particle (charge) and energy currents.
Such a two-parameter CGF is necessary for obtaining later in Sec. \ref{App-eta} the statistics of efficiency in a thermoelectric engine.

We define the particle ($p$) and energy ($e$) current operators
from the rate of change of electron number and electron energy in one of the leads, say $R$, and write
\be
I_p(t)\equiv -  \frac{d N_R^{H}(t)}{dt}, \quad I_e(t) \equiv - \frac{dH_R^{H}(t)}{dt}.
\ee
Here $N_R= \sum_{r \in R}  a_r^{\dagger} a_r$ is the number operator for the
total charge in the right compartment (right lead plus attached acceptor site).
Similarly, $H_R= \sum_{r \in R}  \epsilon_r a_r^{\dagger} a_r$.
The operators are written in the Heisenberg ($H$) picture, thus
they should be evolved with the total Hamiltonian for either model,
$A^H(t)= U^{\dagger}(t)\, A \,U(t)$
where $U(t)= e^{-i H_T t }$.
We follow the convention that the current flowing out of the right lead is positive.
Changes in the total energy and electron number in the $R$ lead during the time interval
$(t_0=0,t)$ ($t_0$ and $t$ are initial and final observation time, respectively), are given by the integrated currents.
\bea
&&Q_{e}(t,t_0)\equiv \int_{t_0=0}^{t}  I_e(t') \, dt'=  H_R(0) -  H_R^{H}(t) ,\nonumber \\
&&Q_{p}(t,t_0)\equiv \int_{t_0=0}^{t}  I_p(t')  \, dt'= N_R(0) -  N_R^{H}(t) .
\eea
Since at any instant  $\big[N_R^H, H_R^H]=0$,
it is possible to construct the so-called ``characteristic  function"
 ${\cal Z}(\lambda_e, \lambda_p)$,
corresponding to the joint probability distribution $P(Q_e,Q_p)$
for the charge and energy currents.
Following the two-time measurement procedure
one can define the characteristic function as
\cite{Esp5-review, Hanggi-review, bijay-ballistic,commM}
\be
{\cal Z}(\lambda_e, \lambda_p) = \Big\langle e^{i \lambda_{e} H_R + i \lambda_{p} N_R} \, e^{-i \lambda_{e} H_R^H(t) - i \lambda_{p} N_R^H(t)} \Big\rangle,
\label{eq:GF}
\ee
where $\lambda_e$ and $\lambda_p$ are the counting fields for energy and particles, respectively.
$\langle \cdots \rangle$ represents an average with respect to the total density matrix
$\rho_T(0)$ at the initial time.
We assume that $\rho_T(0)= \rho_L(0)\otimes \rho_R(0)\otimes \rho_{vib}(0)$,
a factorized-product form for the electronic degrees of freedom and for the
vibrational part.
The leads are maintained in equilibrium at temperature
$T_{\alpha}=1/\beta_{\alpha}$ and chemical potential $\mu_{\alpha}$, $\alpha=L,R$,
and the states are described by the grand canonical distribution function
$\rho_{\alpha}(0)\!=\! \exp[-\beta_{\alpha} (H_{\alpha}-\mu_{\alpha} N_{\alpha})]/Z_{\alpha}$,
with $Z_{\alpha}\!=\!{\rm Tr}\big[\exp[-\beta_{\alpha} (H_{\alpha}-\mu_{\alpha} N_{\alpha})]\big]$ as the
grand canonical partition function.
Eq.~(\ref{eq:GF}) can be reorganized as
\bea
{\cal Z}(\lambda_e, \lambda_p)
&=& {\rm Tr}_T \big[ U_{-\lambda_e/2, -\lambda_p/2}(t)\, \rho_T(0)  \, U^{\dagger}_{\lambda_e/2, \lambda_p/2}(t)\big], \nonumber \\
&=& {\rm Tr}_T \big[ \rho^T_{\lambda_e,\lambda_p}(t)\big], \nonumber \\
&=& {\rm Tr}_{vib} \big[\rho^{vib}_{\lambda_e,\lambda_p}(t)\big].
\label{CGF-master}
\eea
%
The second line introduces the definition of the total, counting-field dependent, density operator.
We trace out its electronic degrees of freedom, (${\rm Tr}_{\rm el}$) and express the characteristic function
in terms of the reduced density matrix $\rho^{vib}_{\lambda_e,\lambda_p}(t)$ for the vibrational mode,
\bea
\rho^{vib}_{\lambda_e,\lambda_p}(t) \equiv
{\rm Tr}_{\rm el} \big[ U_{-\lambda_e/2, -\lambda_p/2}(t)\, \rho_T(0)  \, U^{\dagger}_{\lambda_e/2, \lambda_p/2}(t)\big].
\nonumber\\
\label{eq:vib}
\eea
Note that the forward and backward evolution operators are {\it not} hermitian conjugates. 
For example, the forward propagator is
\bea
&&{U}_{-\lambda_e/2, -\lambda_p/2} (t)=
\nonumber\\
&& \exp\left[{-i \frac{\lambda_e}{2} H_R - i \frac{\lambda_p}{2} N_R}\right]\, {U}(t) \,\exp\left[{i \frac{\lambda_e}{2} H_R + i \frac{\lambda_p}{2} N_R}\right]
\nonumber \\
&&\equiv \exp[-i H_{-\lambda_e/2,-\lambda_p/2} (t)],
\eea
with the counting-field dependent total Hamiltonian
\bea
&&{H}_{-\lambda_e/2,-\lambda_p/2} \equiv
\nonumber\\
&&
\exp\left[{-i \frac{\lambda_e}{2} H_R - i \frac{\lambda_p}{2} N_R}\right]\,H_T \,
 \exp\left[{i \frac{\lambda_e}{2} H_R + i \frac{\lambda_p}{2} N_R}\right] \nonumber \\
&&= H_{\rm el} + H_{vib} + S \otimes \big[g \sum_{l,r} \gamma_{l}^{*} \gamma_r a_{l}^{\dagger} a_{r} e^{\frac{i}{2} ( \lambda_e \epsilon_r + \lambda_p)} + {\rm h.c.} \big].
\nonumber\\
\label{eq:Hlam}
\eea
$S$ is a system operator;
the $H_{DA-AH}$ model is reached when
$S=\sigma_x$, $H_{vib}= \frac{\omega_0}{2} \sigma_z$.
The model $H_{DA-HO}$ is realized with
 $S=(b_0^{\dagger}+b_0)$ and $H_{vib}= \omega_0 b_0^{\dagger} b_0$.
The electron-phonon coupling term here depends on the counting field.
Herein, we use $\lambda$ as a short-hand notation for both $\lambda_e$ and $\lambda_p$.
To facilitate our discussion below, we define the operators $B_{\pm \lambda/2}$ as
\be
B_{\mp\lambda/2}\equiv g \big[ \sum_{l,r} \gamma_{l}^{*} \gamma_r a_{l}^{\dagger} a_{r}  e^{\frac{\pm i}{2} (\epsilon_r {\lambda_e} + {\lambda_p})} + {\rm h.c.}\big].
\label{eq:F}
\ee
%
These operators correspond to the bath operator coupled to the system, see Eqs. (\ref{eq:DA-HO}) and (\ref{eq:DA-AH}),
now dressed by counting fields $\lambda_e, \lambda_p$ as a consequence of the
measurements of charge and energy.
Note that the sign convention for $B$ corresponds to the respective time evolution operator.
The counting-field Hamiltonians [Eq. (\ref{eq:Hlam}) and the complementarity term for the backward evolution]
can be organized in a form convenient
for a perturbation expansion in $g$,
\bea
H_{\pm \lambda/2}&=&H_0+V_{\pm\lambda/2}, 
\nonumber\\
H_0&=&H_{\rm el}+H_{vib},\,\,\,\,\,\ V_{\pm\lambda/2} \equiv S \otimes B_{\pm\lambda/2},
\eea
%
%
with $H_{\rm el}$ incorporating the two metals with the hybridized states, see Eq. (\ref{eq:Hel}).


\subsection{DA-AH model: Quantum master equation approach}
\label{FCS-AH}

We derive the counting field dependent quantum master equation
\cite{Esp5-review, Upendra-heat} for the model (\ref{eq:DA-AH})
under the assumption that the coupling between electron-hole pair generation
 and the vibrational mode is weak  \cite{commentSF}.
Unlike other standard QME approaches for molecular junctions, in which the molecule-metal coupling is
considered weak \cite{Mitra,Esp1}, in the present derivation the metal-molecule hybridization [defined below Eq. (\ref{eq:JLR})]
can be made arbitrarily large, absorbed into the leads spectral density 
by the exact diagonalization procedure presented in Sec. \ref{Model}.
Taking the time-derivative of Eq. (\ref{eq:vib}) we get
\bea
\dot \rho^{vib}_{\lambda}(t) =
{\rm Tr_{\rm el}} \left[ -iV_{-\lambda/2}(t)\rho_{\lambda}^T(t) +i \rho_{\lambda}^T(t)V_{\lambda/2}(t)  \right].
\eea
The operators here are written in the interaction representation, $A(t)=e^{iH_0t}Ae^{-iH_0t}$,
with $H_0$ including the uncoupled electrons and vibration.
By formally integrating this equation we receive the exact form ($\langle B_{\pm\lambda/2}\rangle =0$),
\begin{widetext}
\bea
\dot{\rho}^{{vib}}_{\lambda}(t)  = -\int_{t_0=0}^{t} dt &&'{\rm Tr}_{\rm el} \Big[V_{-\lambda/2}(t) V_{-\lambda/2}(t') \rho^T_{\lambda}(t') + \rho^T_{\lambda}(t') V_{\lambda/2}(t') V_{\lambda/2}(t) \nonumber \\
&& - V_{-\lambda/2}(t') \rho^T_{\lambda}(t') V_{\lambda/2}(t) - V_{-\lambda/2}(t) \rho^T_{\lambda}(t') V_{\lambda/2}(t')\Big].
\label{eq:rhovib}
\eea
\end{widetext}
We now follow standard steps as
in the derivation of the weak coupling-Markov Redfield equation.
The initial condition is assumed fully factorized, $\rho^T_{\lambda}(t')$ is replaced by the initial condition,
$\rho^T_{\lambda}(0)= \rho_T(0)$, and the upper limit of integration in extended to infinity,
assuming Markovianity of the electron baths.
The equation of motion for the counting field dependent
reduced density matrix (describing the dynamics of the vibrational mode) depends on the following
relaxation ($k_d$) and excitation ($k_u$) rates
\bea
k_{d}^{\lambda}
&=& \int_{-\infty}^{\infty} d\tau e^{-i \omega_0 \tau}  \langle B_{\lambda/2}(0)\, B_{-\lambda/2}(\tau) \rangle_{\rm el}, \nonumber \\
k_{u}^{\lambda}&=& \int_{-\infty}^{\infty} d\tau e^{i \omega_0 \tau}
\langle B_{\lambda/2}(0) \, B_{-\lambda/2}(\tau)\rangle_{\rm el}
\nonumber\\
&=& k_{d}^{\lambda} [\omega_0 \to -\omega_0].
\label{eq:kdku}
\eea
Here $\langle \cdots \rangle_{\rm el}= {\rm Tr}_L {\rm Tr}_R [ \cdots \rho_L(0) \rho_R(0)]$.
An explicit calculation of the relaxation rate gives
\begin{widetext}
\bea
k_{d}^{\lambda} &=&
2 \pi g^2 \Big[ \sum_{l,r} |\gamma_l|^2 |\gamma_r|^2 f_L(\epsilon_l) (1-f_R(\epsilon_r))
e^{-i (\lambda_p + \epsilon_r \lambda_e)} \delta(\epsilon_l-\epsilon_r+\omega_0)
\nonumber \\
&+&  \sum_{l,r} |\gamma_l|^2 |\gamma_r|^2 f_R(\epsilon_r) (1-f_L(\epsilon_l)) e^{i (\lambda_p + \epsilon_r \lambda_e)} \delta(\epsilon_l-\epsilon_r-\omega_0)\Big].
\label{eq:rate-down}
\eea
\end{widetext}
The first term represents an inelastic process with an electron hoping from the left lead
to the right one, by absorbing one quanta $\omega_0$, satisfying energy conservation with
${\epsilon}_r = {\epsilon}_l + \omega_0$.
This process, which goes against our convention of a positive current (flowing right to left),
contributes negative charge and energy currents,
reflected by the negative sign for $\lambda_p$ and $\lambda_e$ in the exponent.
The second term in Eq.~(\ref{eq:rate-down}) corresponds to
the reversed process with electron hopping from the right metal to the left, observing
${\epsilon}_l = {\epsilon}_r + \omega_0$, with a positive sign for $\lambda_p$ and $\lambda_e$.
This downward rate assists in cooling the vibrational mode. In the complementary excitation process
electrons lose energy to the vibration, heating up the junction.
Eq. (\ref{eq:rate-down}) can be
decomposed into two separate contributions,
\be
k_{d}^{\lambda}= [k_{d}^{\lambda}]^{L \to R} + [k_{d}^{\lambda}]^{R \to L},
\ee
and similarly for $k_{u}^{\lambda}$.
We define the spectral densities for the baths (metals) as
\be
J_{\alpha}(\epsilon)= 2 \pi g \sum_{k \in \alpha} |\gamma_k|^2 \delta(\epsilon - \epsilon_k).
\label{eq:Jdef}
\ee
Using the transformations (\ref{eq:trans}) we note that these functions are Lorentzian-shaped,
centered around the donor (${\epsilon}_d)$ or acceptor (${\epsilon}_a$) site energies,
\bea
J_L(\epsilon) &=& g \frac{\Gamma_L(\epsilon)}{(\epsilon-\epsilon_d)^2 + \Gamma_L(\epsilon)^2/4},
\nonumber\\
J_R(\epsilon) &=& g \frac{\Gamma_R(\epsilon)}{(\epsilon-\epsilon_a)^2 + \Gamma_R(\epsilon)^2/4},
\label{eq:JLR}
\eea
with $\Gamma_{\alpha}(\epsilon)=2 \pi\sum_{k\in \alpha}v_k^2\delta(\epsilon-\epsilon_k)$.
In terms of these spectral functions, bath induced relaxation rates (\ref{eq:rate-down}) are given by
\bea
&&[k_{d}^{\lambda}]^{L \to R} =
\int \frac{d\epsilon}{2\pi} \Big[ f_L(\epsilon) (1-f_R(\epsilon+\omega_0)) J_L(\epsilon) J_R(\epsilon+\omega_0)
\nonumber\\
&&\times e^{-i(\lambda_p + (\epsilon + \omega_0)\lambda_e)} \Big].
\nonumber\\
&&\big[k_{d}^{\lambda}\big]^{R \to L} =
 \int \frac{d\epsilon}{2\pi} \Big[f_R(\epsilon) (1-f_L(\epsilon+\omega_0)) J_R(\epsilon) J_L(\epsilon+\omega_0)
\nonumber\\
&&\times e^{i(\lambda_p+ \epsilon \lambda_e)}\Big].
\label{eq:rateint}
\eea
%
We also define the  $\lambda=0$ rates from Eqs. (\ref{eq:kdku})-(\ref{eq:rateint}),
only missing the $\lambda$ identifier.
%
The rates are nonzero as long as:
(i) for a given energy, both left and right leads are not fully occupied or empty and
(ii) the overlap between the spectral densities differing by one quanta of energy is non-negligible.
Note that because of the weak electron-phonon coupling approximation, each electron tunnelling process involves
absorption/mission of a single  quanta of energy.
In other words, the dynamics is completely described by single-phonon excitation and relaxation rates.

As mentioned above, we  apply the weak-coupling Born Markov approximation on Eq. (\ref{eq:rhovib}).
By further ignoring off-diagonal coherence elements for the reduced density matrix, we obtain
the population dynamics for the
vibrational states (written here for an arbitrary number of levels, $m=0,1,2,... $),
\bea
\dot{p}^{\lambda}_{m}(t)&=&
 -\big[m k_{d}+(m+1) k_{u}\big]\,{p}^{\lambda}_{m}(t)
\nonumber\\
&+& (m+1) k_{d}^{\lambda} \,{p}^{\lambda}_{m+1}(t) + m k_{u}^{\lambda} \,{p}^{\lambda}_{m-1}(t),
\eea
where ${p}^{\lambda}_m(t)= \langle m|\rho^{vib}_{\lambda}(t)|m\rangle$ and $|m\rangle$ denotes the $m$-th vibrational level.
$k_{d}$ and $k_{u}$ are rates evaluated at $\lambda=0$.
For the DA-AH model, $m=0,1$, it can be shown that off-diagonal coherences do not appear in the Born-Markov approximation
without further assumptions \cite{SegalSF}, and
the dynamics of the population follows
\bea
\dot{p}_0^{\lambda}(t) &=& -k_{u} \, {p}^{\lambda}_{0}(t) +  k_{d}^{\lambda}  \,{p}^{\lambda}_1(t), \nonumber \\
\dot{p}_1^{\lambda}(t) &=&  k_{u}^{\lambda} \, {p}_0^{\lambda}(t)-k_{d} \,{p}_1^{\lambda}(t).
\eea
These equations can be written in a matrix form as
\be
\frac{d|p^{\lambda}(t)\rangle}{dt} = {\cal L}(\lambda) |p^{\lambda}(t)\rangle,
\ee
%
where $|p^{\lambda}(t)\rangle = ({p}^{\lambda}_{0}(t),{p}^{\lambda}_{1}(t))$.
The long-time (steady state) limit defined as
\bea
{\cal G}(\lambda) = \lim_{t \to \infty} \frac{1}{t} \ln {\cal Z} (\lambda)=
\lim_{t \to \infty} \frac{1}{t} \ln \langle I|p^{\lambda}(t)\rangle,
\eea
provides the CGF,
where $\langle I|=(1,1)^T$ is the identity vector. In this limit, only the smallest eigenvalue of the Liouvillian
survives. The final result for the CGF for the DA-AH junction is given by
\bea
{\cal G}_{AH}(\lambda)
=
-\frac{1}{2}(k_{u} + k_{d}) + \frac{1}{2}\sqrt{(k_{u} - k_{d})^2 + 4\, k_{u}^{\lambda} k_{d}^{\lambda}}.
\nonumber\\
\label{eq:CGF-AH}
\eea
We label this CGF by `AH', to highlight the mode anharmonicity. Recall that $\lambda$ collects
two counting fields $\lambda_p$ and $\lambda_e$, for charge and energy, respectively.
The CGF satisfies the fluctuation symmetry
\bea
&&{\cal G}_{\rm AH}(\lambda_e,\lambda_p)
\nonumber\\
&&= {\cal G}_{AH}(-\lambda_e + i(\beta_L-\beta_R), -\lambda_p+i(\beta_R \mu_R-\beta_L \mu_L)),
\nonumber \\
\label{eq:fluc-sym}
\eea
which can be verified by examining the rates
in Eq.~(\ref{eq:rate-down}).  Under the transformations
$\lambda_e \rightarrow -\lambda_e + i(\beta_L-\beta_R)$
and $\lambda_p \rightarrow -\lambda_p+i(\beta_R \mu_R-\beta_L \mu_L)$ the rates transform as
\bea
k_{d}^{\lambda} &\rightarrow& k_{u}^{\lambda} e^{\beta_L \omega_0},
\nonumber \\
k_{u}^{\lambda} &\rightarrow& k_{d}^{\lambda} e^{-\beta_L \omega_0}.
\eea
The extra factors $e^{\pm \beta_L \omega_0}$ cancel out in Eq.~(\ref{eq:CGF-AH}), to satisfy the fluctuation symmetry.

The charge and energy currents and the
corresponding zero frequency noise powers can be readily obtained by taking derivatives of the CGF
with respect to the counting fields.
For example, the particle ($p$) and energy ($e$) currents are obtained from
\bea
\langle I_p\rangle \equiv \frac{\langle Q_p \rangle}{t}
&=& \frac{\partial {\cal G}(\lambda_e,\lambda_p)}{\partial (i \lambda_p)}\Big{|}_{\lambda_e=\lambda_p=0},
\nonumber\\
\langle I_e\rangle \equiv \frac{\langle Q_e \rangle}{t} &=& \frac{\partial {\cal G}(\lambda_e,\lambda_p)}{\partial (i \lambda_e)}\Big{|}_{\lambda_e=\lambda_p=0}.
\label{eq:currents}
\eea
The zero frequency noise of these currents are
\bea
\langle S_p \rangle \equiv \frac{\langle \langle Q_p^2 \rangle \rangle}{t}
&=& \frac{\partial^2 {\cal G}(\lambda_e,\lambda_p)}{\partial (i \lambda_p)^2}\Big{|}_{\lambda_e=\lambda_p=0},
\nonumber\\
\langle S_e \rangle \equiv \frac{\langle \langle Q_e^2 \rangle \rangle}{t}
&=& \frac{\partial^2 {\cal G}(\lambda_e,\lambda_p)}{\partial (i \lambda_e)^2}\Big{|}_{\lambda_e=\lambda_p=0},
\label{eq:noise}
\eea
where
$\langle \langle Q_{e,p}^2 \rangle \rangle = \langle Q_{e,p}^2\rangle - \langle Q_{e,p} \rangle^2$ is the second-cumulant.


\subsection{DA-HO model: Non-equilibrium Green's function approach}
\label{FCS-HO}

We now focus our attention to the DA-HO junction of Fig. \ref{figS}(b), described by
Eq. (\ref{eq:DA-HO}). In this case, electron-hole pair excitation
is coupled to a harmonic oscillator mode.
We employ the NEGF technique \cite{JauhoBook,NEGF-review}
to derive the CGF of this model, again correct up to the second order of the electron-phonon coupling parameter $g$.
Note that the presence of an infinite number of vibrational levels for the HO mode makes it difficult to obtain a closed form for
the CGF under the QME approach, since the Liouvillian is an infinite-dimensional matrix.

\begin{figure}
\includegraphics[width=7cm]{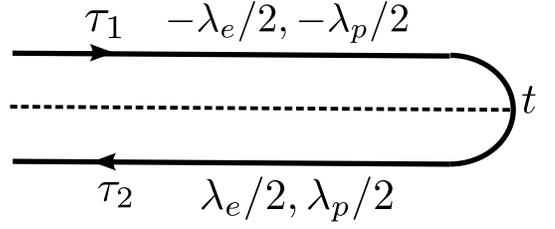}
\caption{Keldysh contour representing the counting statistics problem. The forward upper (+) and backward lower (-) branches
 evolve with different Hamiltonians corresponding to different counting fields. $\tau_1, \tau_2$ are the contour times and $t$ is the final observation time.}
\label{keldysh-contour}
\end{figure}

We begin with Eq.~(\ref{CGF-master}), now identifying the initial time by $t_0$, and
write down the characteristic function on the Keldysh contour (see Fig. \ref{keldysh-contour}) as \cite{commM}
\bea
{\cal Z}(\lambda_e, \lambda_p) &=&
\Big\langle U^{\dagger}_{\lambda_e/2, \lambda_p/2}(t)\,U_{-\lambda_e/2, -\lambda_p/2}(t)\Big\rangle, \nonumber \\
&=& {\rm Tr} \Big[ \rho_T(0) \,T_c\, e^{- i\int_c d\tau {H}^{\lambda(\tau)}_T }\Big].
\label{eq:ZGF}
\eea
The forward upper (backward lower) branch of the Keldysh contour corresponds to the modified unitary evolution $U_{-\lambda_e/2, -\lambda_p/2}(t)$ ($U^{\dagger}_{\lambda_e/2, \lambda_p/2}(t)$).
Evolving the counting fields on two branches of the Keldysh contour with two different signs is the main essence of
counting statistics problems. The normalization condition is trivially satisfied with ${\cal Z}(0,0)=1$. In the above expression
$T_c$ is the contour ordered operator, which orders operators according to their contour time argument;
earliest-time operators appear at the right.
%
$\lambda(\tau)=(\lambda_e(\tau),\lambda_p(\tau))$ is the contour time dependent counting function.
In the upper $(+)$ branch,  $\lambda^{+}(t) =(\lambda_e^+(t),\lambda_p^+(t))= (-\lambda_e/2, -\lambda_p/2)$,
in the lower $(-)$ branch,  $\lambda^{-}(t) = (\lambda_e^-(t),\lambda_p^-(t))=(\lambda_e/2, \lambda_p/2)$.
%
%
Moving to the interaction picture with respect to the Hamiltonian $H_0=H_{\rm el}+ H_{vib}$,
we write the characteristic function as
\be
{\cal Z}(\lambda_e, \lambda_p)= \big \langle T_c \exp(-i \int_{c}\,d\tau \,{V}^{\lambda(\tau)}(\tau) )\big \rangle,
\label{eq-Z-start}
\ee
where the counting field dependent interaction term is
\bea
V^{\lambda(\tau)}(\tau)&=&\big[b_0(\tau) + b_0^{\dagger}(\tau)\big]
\nonumber\\
&\times&g \sum_{l,r} \gamma_l^* \gamma_r \,
e^{ -i (\lambda_p(\tau) + \epsilon_r \lambda_e(\tau))} \,
a_l^{\dagger}(\tau) \, a_r(\tau) \, + {\rm h.c.}
\nonumber \\
&=& g \sum_{l,r} \gamma_l^* \gamma_r  \, a_l^{\dagger}(\tau) \,\tilde{a}_r(\tau) \,
\big[b_0(\tau) + b_0^{\dagger}(\tau)\big] + {\rm h.c.}
\nonumber
\eea
introducing the short notation $\tilde{a}_r(\tau)\equiv e^{ -i (\lambda_p(\tau) + \epsilon_r \lambda_e(\tau))} a_r(\tau)$.
Our objective is to  calculate the CGF, correct up to second-order in electron-phonon coupling
but exact to all orders in the metal-molecule hybridization.
It can be shown that a naive perturbative calculation of Eq.~(\ref{eq-Z-start})
in terms of the electron-phonon coupling $g$ leads to a violation of the non-equilibrium fluctuation symmetry.
Moreover,  such a perturbative treatment cannot capture the correct non-equilibrium phonon distribution,
and it will lead to (an incorrect) long-time solution for the vibrational density matrix which depends on
the initial- arbitrary state $\rho_{vib}(0)$.

\begin{figure} [htpb]
\includegraphics[width=8cm]{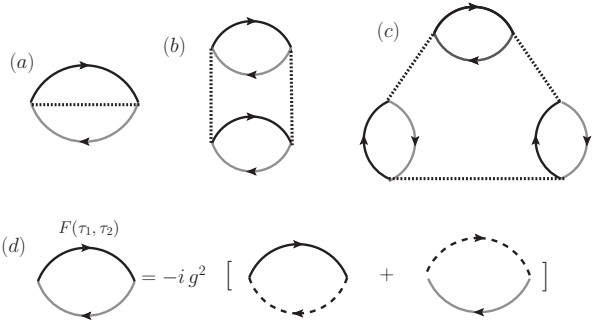}
\caption{Ring type Feynman diagrams in contour time.
(a) Second-order, (b) fourth-order, and (c) sixth-order diagrams in the electron-phonon coupling.
The dotted line represents the phonon Green's function $D_0$. Closed
loops are the electron-hole propagator ${F}(\tau_1,\tau_2)$, the sum
of two diagrams (d) consisting of the bare left (solid) and right (dashed)  leads
Green's functions.}
\label{Feynman_diagrams}
\end{figure}

In order to restore the fluctuation symmetry, and obtain the correct non-equilibrium phonon distribution
[which depends on the temperatures of the electronic baths and their chemical potentials, but not on 
$\rho_{vib}(0)$],
one has to sum over an infinite subclass of diagrams in this perturbative expansion.
This procedure takes into account all electron scattering processes which are facilitated by
the absorption or emission of a {\it single} quanta $\omega_0$. Physically, the summation collects
not only sequentially tunneling electrons, but all coordinated multi-tunneling processes,
albeit with each electron interacting with the mode to the lowest order, to absorb/emit each a single quanta $\omega_0$.
This summation can be achieved by exploiting the random-phase approximation (RPA) as done in Refs.
\cite{Utsumi1,Utsumi2, Atland}, also referred to as the self-consistent Born approximation \cite{SCBA}. 
Summing over a particular set of  diagrams (ring type) in the perturbative series,
see Fig.~\ref{Feynman_diagrams}, we reach the following expression, 
\bea
\ln {\cal Z}_{\rm RPA}(\lambda_e,\lambda_p)
= -\frac{1}{2}\, {\rm Tr}_{\tau} \,\ln \big[I- D_0(\tau,\tau') \tilde{F}(\tau',\tau)\big].
\nonumber\\
\label{CGF-RPA}
\eea
The symbol ${\rm Tr}_{\tau}$ denotes an integration over contour time variables $(\tau,\tau')$.
For example,
\be
{\rm Tr}_{\tau} \big[D_0(\tau,\tau') \tilde{F}(\tau',\tau)\big] = \int d\tau \int d\tau' D_0(\tau,\tau') \tilde{F}(\tau',\tau).
\ee
Here, $I$ is the identity matrix in the Keldysh space and $D_0(\tau,\tau')$ is the free phonon Green's function,
\bea
D_{0}(\tau_1,\tau_2)= -i \langle T_c X(\tau_1) X(\tau_2)\rangle,
\eea
with $X\equiv (b_0+b_0^{\dagger})$, proportional to the phonon displacement operator.
$\tilde{F}(\tau,\tau')$ is the counting-field dependent electron-hole Green's function. It
describes electron hopping processes from the left to the right lead, and vice versa,
\bea
\tilde{F}(\tau_1,\tau_2) &=&
-i g^2 \,\sum_{l\in L,r \in R} |\gamma_l|^2 \gamma_r|^2
\nonumber\\
&\times&
\big[ g_l (\tau_1,\tau_2) \tilde{g}_r(\tau_2,\tau_1) + \tilde{g}_r(\tau_1,\tau_2) g_l(\tau_2,\tau_1)\big].
\nonumber\\
\label{eh-prop}
\eea
This expression is symmetric under the exchange of the contour time parameters $\tau_1$ and $\tau_2$.
Recall that the tilde symbol advices that the Green's function is $\lambda$ dependent.
This propagator involves free electron Green's functions for the left and right leads,
\bea
g_{l}(\tau_1,\tau_2)&=& - i \,\langle T_c a_{l}(\tau_1) a_{l}^{\dagger}(\tau_2)\rangle, \quad
\nonumber\\
\tilde{g}_{r}(\tau_1,\tau_2)&=& - i \langle T_c \tilde{a}_{r}(\tau_1) \tilde{a}_{r}^{\dagger}(\tau_2)\rangle.
\eea
Explicit expressions for different components of these Green's functions in real time
are given in Appendix A. Here, we write down the lesser component, given as,
\begin{widetext}
\bea
\tilde{F}^{<} (\omega) &=& -i \, 2 \pi g^2 \Big[ \sum_{l,r} |\gamma_l|^2 |\gamma_r|^2 f_L(\epsilon_l) (1-f_R(\epsilon_r)) e^{-i(\lambda_p+\epsilon_r \lambda_e)} \delta(\epsilon_l-\epsilon_r-\omega)
\nonumber \\
&+&  \sum_{l,r} |\gamma_l|^2 |\gamma_r|^2 f_R(\epsilon_r) (1-f_L(\epsilon_l)) e^{i(\lambda_p+\epsilon_r\lambda_e)} \delta(\epsilon_l-\epsilon_r+\omega)\Big].
\eea
\end{widetext}
The greater component is obtained from  $\tilde{F}^>(\omega)=\tilde{F}^<(-\omega)$.
It is  clear that the greater (lesser) component corresponds to the electronic bath-induced transition
rates within the vibrational mode,
$k_d^{\lambda}$ ($k_u^{\lambda}$), see definitions (\ref{eq:kdku}) and a physical explanation below Eq. (\ref{eq:rate-down}).

%


Projecting Eq.~(\ref{CGF-RPA}) to the real time and invoking the steady state limit by taking $t_0 \to -\infty$,
we write the CGF as \cite{NEGF-review,bijay-ballistic},  
\bea
{\cal G}(\lambda_e,\lambda_p)&=&  
\lim_{t \to \infty} \frac{1}{t}
\ln {\cal Z}_{\rm RPA}(\lambda_e,\lambda_p)
\nonumber\\
&=&
- \int \frac{d\omega}{4\pi} \ln \det D_{\lambda}^{-1}(\omega)
\label{CGF-H-first}
\eea
where $D_{\lambda}^{-1}(\omega)= D_0^{-1}(\omega)- \sigma_z \,\tilde{F}(\omega) \,\sigma_z$ with
 $\sigma_z$ as the third Pauli matrix.
Note that we renormalized the CGF with a counting field independent term
$D_0^{-1}$. The matrix $D_{\lambda}^{-1}$ can be written explicitly as
\bea
D_{\lambda}^{-1}(\omega)=
\left( \begin{array}{cc}
                         [D_0^r]^{-1}(\omega)-F^t(\omega) & \tilde{F}^{<}(\omega)   \\
                          \tilde{F}^{>}(\omega) & -[D_0^a]^{-1}(\omega)-F^{\bar{t}}(\omega)
\end{array}\right)
\nonumber
\eea
where $r,a, t, \bar{t}, <, >$ stand for the retarded, advanced, time-ordered, anti-time ordered, lesser and greater components of the Green's functions, respectively.
The free-phonon Green's functions $D_0^{r,a}(\omega)$ are given by
\be
D_0^{r}(\omega) = \frac{2 \omega_0}{(\omega+i\eta)^2 -\omega_0^2}, \quad  D_0^{a}(\omega)= [D_0^r(\omega)] ^*,
\ee
where $\eta$ is an infinitesimal positive number, introduced to preserve the causality of the retarded Green's function.
The determinant of ${D}_{\lambda}^{-1}(\omega)$ can be immediately evaluated,
\bea
\det{D}_{\lambda}^{-1}(\omega)= - \Big[\frac{\omega^2-\omega_0^2}{2 \omega_0} -\Big(\frac{F^t\!-\!F^{\bar{t}}}{2}\Big)\Big]^2- A_{\lambda}(\omega).
\nonumber\\
\eea
Here
\be
A_{\lambda}(\omega)\equiv \tilde{F}^<(\omega)\tilde{F}^>(\omega) - \frac{\big(F^<(\omega)\!+\! F^{>}(\omega)\big)^2}{4},
\ee
is written solely in terms of the electron-hole Green's function, and it will end
up being the central quantity in this problem. The above determinant was simplified using the
 identity $F^t(\omega)+ F^{\bar{t}}(\omega)\!=\! F^>(\omega) + F^<(\omega)$,
see Appendix A.
We now take the $\lambda$ derivative of the CGF in Eq.~(\ref{CGF-H-first}), to obtain
\be
\partial_{\lambda} {\cal G}(\lambda)=-\int \frac{d\omega}{4 \pi} \frac{\partial_{\lambda} A_{\lambda}(\omega)}{ \Big[\frac{\omega^2-\omega_0^2}{2 \omega_0} -\Big(\frac{F^t-F^{\bar{t}}}{2}\Big)\Big]^2+A_{\lambda}(\omega)}.
\label{eq:RPA2}
\ee
The integration can be performed to the leading order in the electron-phonon coupling $g$.
To the lowest nontrivial order in the electron-phonon coupling
the location of the poles can be approximated by
\bea
\pm \left\{\omega_0 + {\rm Re}[F^r(\omega_0)]\pm i \sqrt{A_{\lambda}(\omega_0)} \right\},
\eea
where we identify ${\rm Re}[F^r]=(F^t-F^{\bar{t}})/2$.
Employing the residue theorem, the integration in Eq. (\ref{eq:RPA2}) can be performed, resulting in
\bea
\partial_{\lambda} {\cal G}(\lambda) \approx -\partial_{\lambda} \sqrt{A_{\lambda}(\omega_0)},
\eea
We now formally identify the lesser and greater components of the electron-hole Green's functions
in $A_{\lambda}(\omega_0)$ with the excitation and relaxation rates, defined in the QME approach in Sec. \ref{FCS-AH},
see  Eq.~(\ref{eq:rate-down}).  They are given by $F^>(\omega_0)= -i k_{d}$ and
$F^<(\omega_0)= -i k_{u}$. In the presence of counting fields,  $\tilde{F}^>(\omega_0)= -i k^{\lambda}_{d}$,
$\tilde{F}^<(\omega_0)= -i k^{\lambda}_{u}$. The final expression for the CGF is
\bea
{\cal G}_{HO}(\lambda) = \frac{1}{2}(k_{d} - k_{u})
- \frac{1}{2}\sqrt{(k_{u} + k_{d})^2 - 4 k_{u}^{\lambda} k_{d}^{\lambda}}.
\nonumber\\
\label{eq:CGF-H}
\eea
Remarkably, the expressions for ${\cal G}_{HO}(\lambda_e,\lambda_p)$ and ${\cal G}_{AH}(\lambda_e,\lambda_p)$ in
Eq.~(\ref{eq:CGF-AH}) are very similar, besides the sign differences, though  they were obtained
via two completely different approaches. The sign difference reflects the different normalization,
with the AH mode conserving population in the two states of the mode, while in the harmonic mode
all levels are occupied at nonzero temperature.
The RPA approximation restores the fluctuation symmetry, Eq. (\ref{eq:fluc-sym}).

The CGFs for the DA-AH and DA-HO models, Eqs. (\ref{eq:CGF-AH}) and (\ref{eq:CGF-H}), respectively,
are the main analytical results of this paper \cite{commentSF}.
It should be emphasized that only few impurity models, essentially, variants
of the single dot Anderson model \cite{gogolin,Fazio,Lei,Levy1,Utsumi1,gernot},
can be solved analytically, within certain approximations, to provide
the CGF and expose charge and energy statistics under interactions.
Our work here substantially extends these efforts, by solving the FCS of
a vibrationally assisted two-site electronic conduction.
In Sec. \ref{App-IV} we derive and simulate charge and energy currents and their noise using
 Eqs. (\ref{eq:currents}) and (\ref{eq:noise}).
As a further nontrivial application, we employ the CGFs in Sec. \ref{App-eta} to simulate
fluctuations of the thermoelectric efficiency.

\begin{figure*}
\centering
\includegraphics[width=12cm]{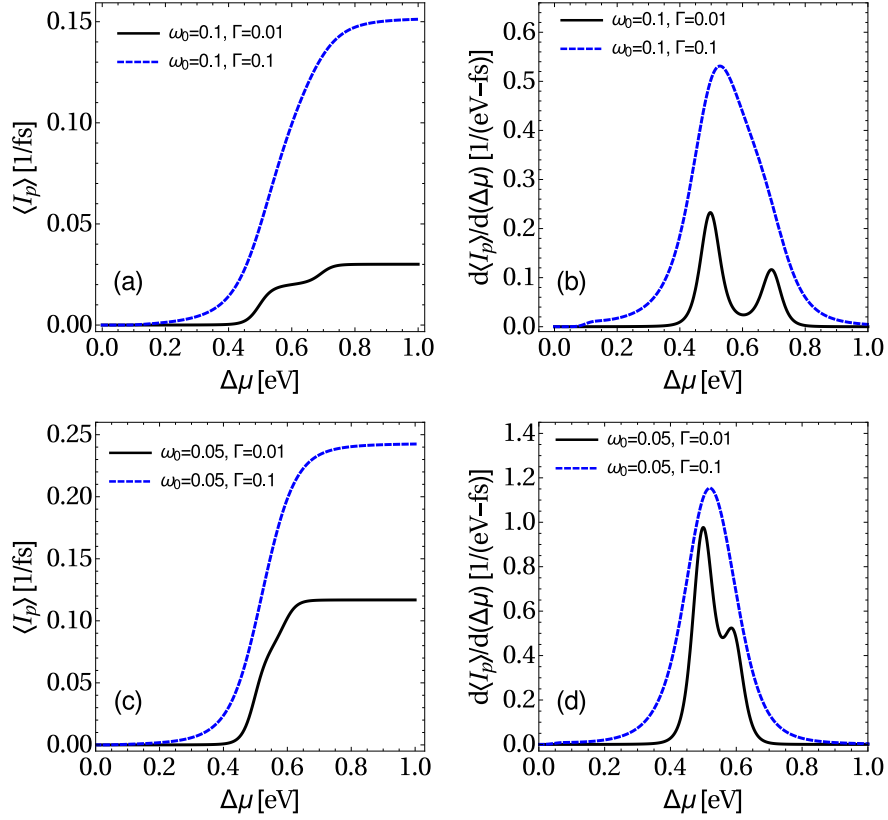}
\caption{(color online)
Charge current and differential conductance as a function of voltage bias for the DA-AH model
with (a)-(b) $\omega_0=0.1$ eV, and (c)-(d) $\omega_0=0.05$ eV.
Other parameters are
$\epsilon_d\!=\!\epsilon_a\!=\!0.25$ eV, $T=T_L\!=\!T_R\!=\!100$ K,
$g=0.1$ eV, $\Gamma=\!0.01$ eV (solid),
$\Gamma=0.1$ eV (dashed).
}
\label{current-AH}
\end{figure*}

\begin{figure*}
\includegraphics[width=12cm]{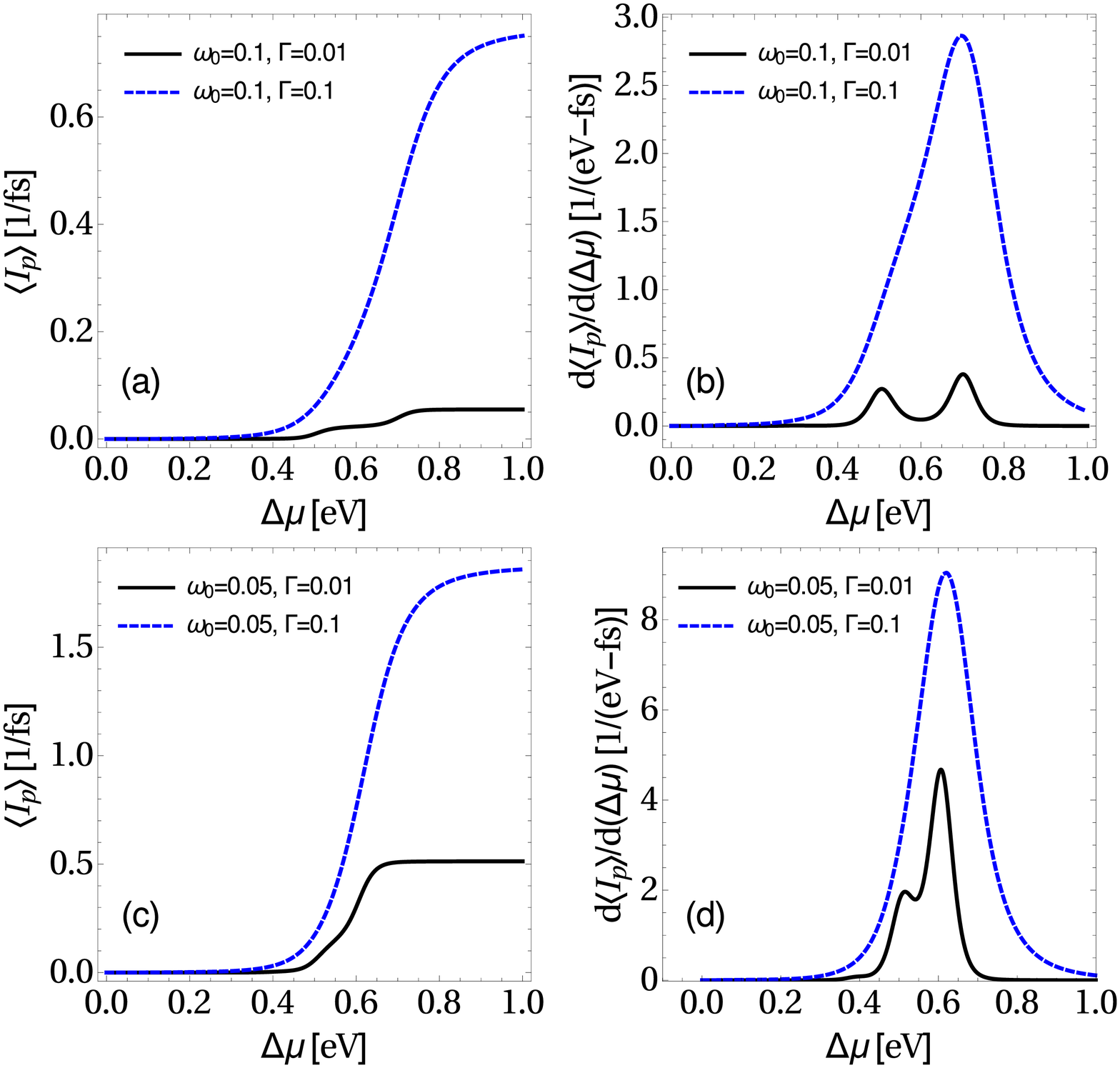}
\caption{(color online)
Charge current and differential conductance as a function of voltage bias for the DA-HO model
with (a)-(b) $\omega_0=0.1$ eV and (c)-(d) $\omega_0=0.05$ eV.
Parameters are the same as in Fig.~\ref{current-AH},
but the mode is further coupled to a dissipative phonon bath with
$\Gamma_{\rm ph}=0.05$ eV and $T_{\rm ph}=T_L=T_R$.}
\label{current-H}
\end{figure*}


\section{Applications}
\label{Applications}

\subsection{Charge current and Fano factor far from equilibrium}
\label{App-IV}

We present here simulations for the charge current and its noise in the DA junction. Particularly, we
examine signatures of mode harmonicity in transport.
In Ref. \cite{bijay_thermo} we further explore  fingerprints of vibrational anharmonicities
in linear response quantities: the electronic thermal conductance, the thermopower, and the thermoelectric figure of merit.

We obtain closed-form expressions for the currents and high order cumulants
by taking partial derivatives of the CGF with respect to the counting fields, see
Eqs. (\ref{eq:currents}) and (\ref{eq:noise}).
The particle ($p$) and energy ($e$) currents in the DA-AH and DA-HO models are given by
\bea
\langle I_p^{AH/HO} \rangle &=& 2 \, \frac{k_{d} ^{R \to L} k_{u}^{R \to L} - k_{d}^{L \to R} k_{u}^{L \to R}}{k_{d}
+ s k_{u}}, \nonumber \\
\langle I_e^{AH/HO} \rangle &=& \frac{k_{d}\,\big[ \partial_{(i \lambda_e)}
k_{u}^{\lambda}|_{\lambda=0}\big]+ k_{u} \big[ \partial_{(i \lambda_e)} k_{d}^{\lambda}|_{\lambda=0}\big]}{k_{d} + s \, k_{u}},
\nonumber\\
\label{eq:AH/HO-current}
\eea
%
where $s =+1$ ($s=-1)$ for the AH (HO) mode.
Note that we did not simplify the expression for the energy current above; the derivatives
return energy transfer rates which are analogues
to Eq.  (\ref{eq:rateint}), only with an additional energy variable in the integrand.
The average  heat current, extracted from the right terminal, is defined as
$\langle I_q^{AH/HO}\rangle = \langle I_e^{AH/HO} \rangle-\mu_R \langle I_p^{AH/HO} \rangle$.
We further write down closed expressions for the particle-current noise,
\bea
\langle S_p^{AH/HO}\rangle&=& -\frac{2 s}{k_d+s k_u}\langle I_p^{AH/HO}\rangle^2
\nonumber\\
&+& \frac{4}{k_d+s k_u} \left( k_u^{L \to R} k_d^{L\to R} + k_u ^{R\to L}k_d^{R \to L} \right).
\nonumber\\
\label{eq:AH/HO-noise}
\eea
The first term hands over a strictly non-equilibrium (finite-bias) noise.
The second term survives even when the bias voltage is zero,
thus we refer to it as the equilibrium contribution
(though it is somewhat modified with bias).
At low bias and low temperatures, $k_u \ll k_d$, thus $\langle S_p^{AH}\rangle \sim
 \langle S_p^{HO}\rangle$.
At finite bias, significant differences show up, as we discuss in the text following Figs. \ref{fano-AH} and
\ref{fano-H}.

The Fano factor, defined as  the ratio of the noise to the current,
$F\equiv \langle S_p\rangle /  \langle I_p\rangle$, receives a rather transparent form
\bea
F^{AH/HO}&=&-\frac{2s\langle I_p^{AH,HO} \rangle }{k_d+sk_u}
\nonumber\\
&+& 2\frac
{k_d^{R \to L} k_u^{R \to L} + k_d^{L \to R} k_u^{L \to R} } {k_d^{R \to L} k_u^{R \to L} - k_d^{L \to R} k_u^{L \to R}}.
\label{eq:F}
\eea
The second term here does not depend on the mode harmonicity/anharmonicity: At finite bias
$k_{d,u}^{R\to L} > k_{d,u}^{L\to R}$, thus this term roughly takes on the value 2, besides
at asymptotically small biases when the denominator drops to zero since the current itself is diminishing. 
The first term in Eq. (\ref{eq:F}), in contrast, depends on the mode harmonicity,
it strongly varies with the bias voltage, and physically it corresponds to
the ratio between two rates: charge transfer through the junction and transitions between vibrational states 
within the attached mode.

Before presenting results at finite voltage and temperature, we derive
scaling relations for the current and its noise in the large voltage - zero temperature
limit, when metal-molecule hybridization is large, $J_L(\omega)\rightarrow \frac{4g}{\Gamma_L}$, 
$J_R(\omega)\rightarrow \frac{4g}{\Gamma_R}$, see Eq. (\ref{eq:JLR}).
This limit will allow us to pinpoint on fundamental differences between the HO and the AH mode models.
As well, scaling relations will be contrasted with results from the Anderson-Holstein model.
We introduce the notation $\bar g^2\equiv8g^2/\pi$, 
assume zero electronic temperature and a large voltage bias, $\Delta\mu =\mu_R-\mu_L\gg \epsilon_{d,a},
\Gamma_{L,R}, \omega_0$,
and obtain from   Eq. (\ref{eq:rateint}) the $\lambda=0$ rates,
\bea
k_d&\approx&k_d^{R \to L}\approx \frac{\bar{g}^2 \omega_0} {\Gamma_L \Gamma_R} \Big(\frac{\Delta \mu}{\omega_0}+1\Big), \nonumber \\
k_u&\approx&k_u^{R \to L}\approx \frac{\bar{g}^2 \omega_0} {\Gamma_L \Gamma_R} \Big(\frac{\Delta \mu}{\omega_0}-1\Big),
\eea
with negligible left-to-right rate constants. 
Eqs. (\ref{eq:AH/HO-current})-(\ref{eq:AH/HO-noise}) for the particle current then reduce to
\bea
\langle I_p^{HO} \rangle &=&\frac{\bar g^2\omega_0}{\Gamma_L\Gamma_R} \left( \frac{\Delta \mu^2}{\omega_0^2}-1\right),
\nonumber\\
\langle I_p^{AH} \rangle &=& \frac{\bar g^2\Delta \mu}{\Gamma_L\Gamma_R}\left( 1- \frac{\omega_0^2}{\Delta\mu^2}\right).
\eea
Similarly, we derive the particle current noise
\bea
\langle S_p^{HO} \rangle &=&\frac{\bar g^2\omega_0}{\Gamma_L\Gamma_R} \left( \frac{\Delta \mu^4}{\omega_0^4}-1\right),
\nonumber\\
\langle S_p^{AH} \rangle &=& \frac{\bar g^2\Delta \mu}{\Gamma_L\Gamma_R}\left( 1- \frac{\omega_0^4}{\Delta\mu^4}\right),
\eea
and the Fano factor
\bea
F^{HO}&=& \frac{\Delta\mu^2}{\omega_0^2} +1 \sim \frac{\Delta\mu^2}{\omega_0^2},
\nonumber\\
F^{AH} &=& 1+\frac{\omega_0^2}{\Delta\mu^2} \sim 1.
\eea
Thus, while at low bias and low temperature the DA-AH and DA-HO models similarly behave, 
at high voltage fundamental differences are displayed, particularly in the current statistics,
(see e.g. Fig. \ref{fano-AH}).
It can be further proved that in the DA-AH model higher order cumulants scale as $C_{n+1}/C_n\propto 1$,
while the DA-HO model supports $C_{n+1}/C_n\propto \Delta\mu^2/\omega_0^2$.
In contrast, the single impurity Anderson Holstein model shows a {\it different} scaling altogether,
$C_{n+1}/C_n\propto \Delta\mu/\omega_0$ \cite{Levy2}.
These three models thus display distinct noise characteristics, a useful input for identifying the nature of 
electron-phonon coupling in conducting junctions.

We proceed by presenting simulation results at finite temperature and bias, 
focusing on the large bias limit rather than the linear response behavior.
We set the Fermi energy $\mu$ at zero, and adjust the chemical potentials
of the leads symmetrically around the Fermi energy, $\mu_L= -\mu_R$, $\Delta \mu= \mu_R-\mu_L$.
For simplicity, we align the donor and acceptor energies at the same value and use
${\epsilon}_d\!=\!{\epsilon}_a\!=\!{\epsilon}_0=0.25$ eV.
The junction is assumed symmetric with $\Gamma\equiv\Gamma_{L,R}$, and
the temperature is taken rather low,
$T<\omega_0,\Gamma,\epsilon_0$.
We employ $g=0.1$ eV for the electron-vibration coupling energy;
this value may seem large given the perturbative nature of our treatment, requiring
$g/\omega_0\ll1$. However,  since in the present weak-coupling limit the current
simply scales as $g^2$, our simulations below are representative
and can be immediately translated to consider other values for $g$ \cite{comment2}.
Simulations were performed by evaluating numerically the rates, assuming metals with
a wide bandwidth $D$ (larger than all other energy scales), and an energy-independent hybridization $\Gamma$.

In Fig.~\ref{current-AH} we study the DA-AH model and
display the current and its derivative, the differential conductance,
as a function of the applied voltage bias.
Panels (a) and (b) illustrate results with a relatively high mode frequency,
$\omega_0=0.1$ eV, using $\Gamma=0.01$ eV or $0.1$ eV and $T$=100 K.
We find that when the hybridization energy is small, $\Gamma\ll\omega_0$,
the current increases in two steps positioned at
$\Delta \mu \approx 2 \epsilon_0$ and $\Delta \mu \approx 2 (\epsilon_0 + \omega_0)$.
These steps are clearly resolved as a two-peak structure in the differential conductance,
see Fig.~\ref{current-AH}(b).

The location of these peaks can be reasoned by
investigating the expression for the current, Eq.~(\ref{eq:AH/HO-current}):
For the given parameters with site energy $\epsilon_0 >\mu$ and low temperatures,
the rates $k_{u}^{R \to L}$ and $k_{d}^{R \to L}$ dominate the current at finite bias
whereas $k_{u}^{L \to R}$ and $k_{d}^{L \to R}$ are negligible.
As we gradually raise the bias, we find that the vibrational relaxation rate $k_d^{R \to L}$ significantly increases
once $\mu_R$ approaches the site energy, $\mu_R = \Delta \mu/2 \approx {\epsilon}_0$,
as the chemical potential precisely sits then within a region of a high molecular electronic density of states,
reflected by the first jump in the current.
At further higher biases, $\mu_R \approx {\epsilon}_0 + \omega_0$,
the rate $k_u^{R \to L}$ is now strongly enhanced since  an excess energy $\omega_0$ is available for mode excitation,
producing the second peak.
The energy gap between the peaks is therefore given by $2\omega_0$.
At high hybridization energies, $\Gamma \gtrsim \omega_0$, the current reaches
higher values in comparison to the weak hybridization case, and it increases monotonically before saturation.
However, in this large $\Gamma$ limit  the differential conductance
reveals only a single broad peak centered around $\Delta \mu\sim 2 {\epsilon}_0$.


Panels (c)-(d) in Fig. \ref{current-AH} illustrate the behavior of the current and the differential conductance
when adopting a smaller value for the vibrational frequency, $\omega_0=0.05$ eV.
We observe similar features as in the previous larger-$\omega_0$ case, only
the broadening $\Gamma$ now conceals the two-peak structure.
It is also notable that the first resonance peak at $\Delta \mu =2\epsilon_0$ is higher in magnitude than
the second jump at $\Delta \mu =2(\epsilon_0+\omega_0)$.

Transport in the  DA-HO junction is similarly examined in Fig. \ref{current-H}.
The magnitude of the current is higher in comparison to the DA-AH case (Fig. \ref{current-AH}),
for both weak and strong hybridization energies, due to the availability of
many additional channels for excitations and relaxations, once the bias exceeds the value
$2(\epsilon_0+\omega_0)$.
In fact, at high bias voltage the charge current diverges and the vibrational mode becomes overly-heated,
a phenomena referred to as ``vibrational instability" \cite{Brandbyge}.
This heating effect can be controlled and avoided if dissipation of energy from the single-molecular vibration to
an additional bath (besides the metals) is  allowed \cite{SegalSF,SegalINFPI}, see Appendix B for details.
Mathematically, this divergence is reflected by the denominator in the expression for the current:
At large bias the current-induced excitation rate exceeds the relaxation rate
(as we break detailed balance at finite bias) \cite{Brandbyge}.
To remove the vibrational instability, results in Fig.~\ref{current-H} were obtained by
attaching the molecular mode to a secondary phonon bath 
with a finite phonon damping rate $\Gamma_{\rm ph} =0.05$ eV and $T_{\rm ph}=T_L=T_R$.
Interestingly, inspecting the differential conductance, we observe that 
the peak at $\Delta \mu = 2({\epsilon}_0+ \omega_0)$ is more pronounced relative to
the first peak at $\Delta \mu = 2{\epsilon}_0$. This trend is opposite to
the DA-AH case in Fig. \ref{current-AH}. It is explained by noting that 
the second peak corresponds to the opening up of many channels for charge transfer in the case of a harmonic mode.

\begin{figure}
\includegraphics[width=8cm]{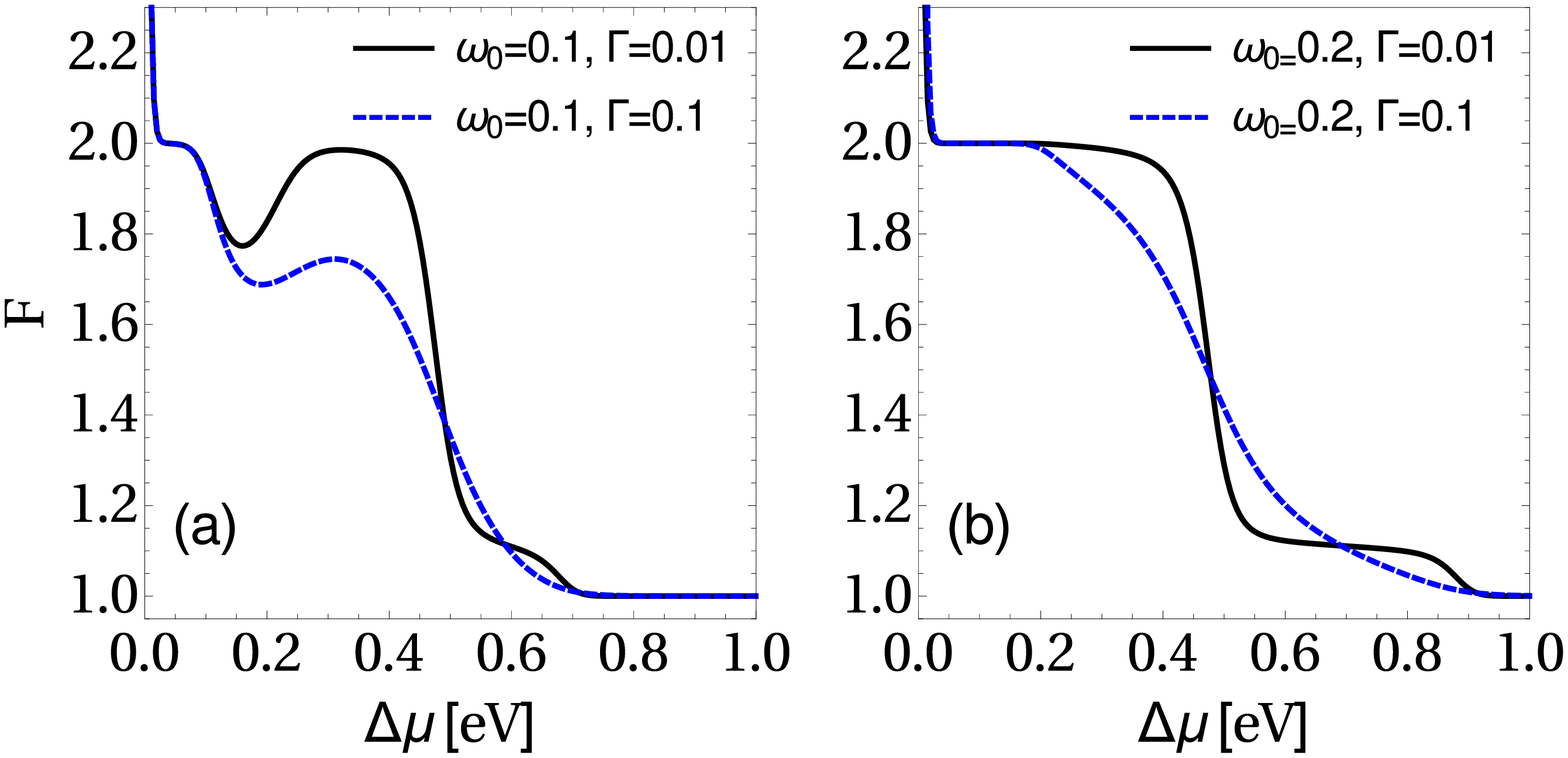}
\caption{(color online)
Fano factor as a function of voltage bias for the DA-AH model.
Junction's parameters are same as in Fig. \ref{current-AH}.}
\label{fano-AH}
\end{figure}

\begin{figure}
\includegraphics[width=8cm]{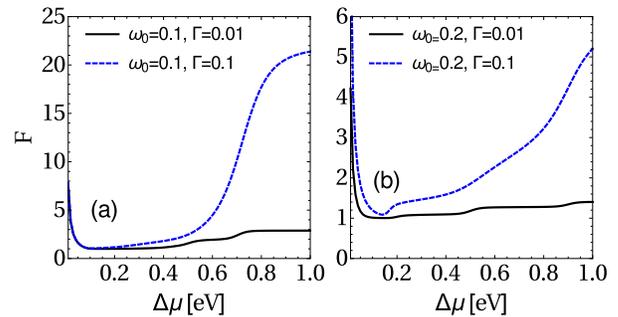}
\caption{(color online)
Fano factor as a function of applied voltage bias for the DA-HO model,
with parameters as in Fig. \ref{current-H}.}
\label{fano-H}
\end{figure}

The zero-frequency Fano factor, Eq. (\ref{eq:F}),
is investigated in Figs. \ref{fano-AH} and \ref{fano-H}.
We find that this measure strongly reflects the nature of the vibrational mode:
In the DA-AH model the Fano factor shows a super-Poissonian behavior at low-intermediate biases, but
in the high bias limit $\Delta \mu> 2(\epsilon_0 + \omega_0)$ it reaches the value 1,
reflecting a Poissonian behavior.
In this high bias limit, we receive analytically $F=-1+2$, where the first (second) term
in Eq.~(\ref{eq:AH/HO-noise}) contributes $-1$ ($2$).
While we cannot offer a fundamental understanding of the involved features in Figs.  \ref{fano-AH},
we confirm that they emerge from the behavior of the first term in Eq. (\ref{eq:F}),
while the second-equilibrium term maintains the value $\sim $2 at the relevant range of applied voltage.
The DA-HO junction shows a very different behavior, see
Fig. \ref{fano-H}. Here, a Poissonian behavior takes place at relatively low biases,
$\Delta \mu < \epsilon_0$, but beyond that $F$ is always super-Poissonian,
reaching high values when many vibrational states
participate in the conductance.

Other theoretical studies have confirmed that
molecular junctions may reach very high noise levels due to electron-vibration scattering processes
\cite{Koch1,Koch2, Novotni, Thoss1}.
It should be emphasized however that in these calculations large $F$ values were materialized
under the assumption of a strong electron-phonon interaction, while
the molecule-lead coupling was assumed weak.
In contrast, we are concerned here with precisely the opposite arrangement:
weak electron-phonon interaction but arbitrary large metal-molecule coupling, and
we reach large values for $F$ due to the breakdown of
the detailed balance relation by the applied bias voltage leading to the participation of
many vibrational states in transport \cite{Brandbyge,SegalSF}.

To summarize our observations in this Section,
the current and  its noise can reveal information on the vibrational mode participating in the transport
process, as well as provide input on the hybridization strength of the molecule to the leads.
The differential conductance shows a two-peak structure.
The separation between the peaks corresponds to (twice) the vibrational frequency, and
the relative peaks' height can be attributed to mode harmonicity.
A strong hybridization, $\Gamma\gtrsim \omega_0$, smears out the double-peak structure to form a single-asymmetric feature.
Genuine anharmonicity, e.g., in the form of a morse potential rather than a two-state system,  should lead to a
differential conductance similar
to that obtained in the harmonic case, as long as $\Gamma$ is greater than the
anharmonic energy scale ($\Gamma>\omega_0^2/D_e$, with $D_e$ as the dissociation energy in the morse potential).
The significant qualitative differences in the noise characteristics between the DA-AH and the DA-HO junctions
could assist in identifying the participating ``impurity" mode.


\subsection{Thermoelectric efficiency and its statistics}
\label{App-eta}

\subsubsection{Large deviation function for efficiency}
\label{etaLDF}

In this Section we study the operation of the DA junction as a thermoelectric engine.
We explore the device averaged efficiency under certain conditions
and the statistics of efficiency fluctuations, which should play an important role in small devices as opposed to the bulk.
In a recent study, Esposito et al. had analyzed the thermoelectric efficiency statistics in a purely coherent
charge transport model \cite{Esp1}. 
Classical models were examined in other studies \cite{Esp4}.
The DA junction offers a rich opportunity to examine the thermoelectric efficiency beyond linear response,
explore the new concept of efficiency fluctuations,
and interrogate the role of quantum effects and many-body interactions on the operation of a molecular thermoelectric engine.
The DA junction is particularly interesting in this context:  As exemplified in Ref. \cite{bijay_thermo} and
below in Fig. \ref{eff-power}(b), the macroscopic thermoelectric efficiency is
{\it  identical} in the DA-AH and DA-HO models; only fluctuations of efficiency reveal signatures of molecular anharmonicity.

To operate the device as a thermoelectric engine we
set $T_L < T_R$ and $\mu_L > \mu_R$.
The macroscopic thermoelectric (TE) efficiency $\bar\eta_{TE}$ is defined as the ratio between the averaged power generated
by the engine,
\bea
-\dot W\equiv(\mu_L-\mu_R)\langle I_p\rangle
\eea
to the heat absorbed from the hot reservoir,
\bea
\dot Q\equiv \langle I_q\rangle= \langle I_e\rangle - \mu_R \langle I_p\rangle.
\eea
Namely, $\bar\eta_{TE}=\frac{(\mu_L-\mu_R)\langle I_p\rangle}{\langle I_q\rangle }$.
According to the second law, the engine's efficiency is upper bounded, ${\bar \eta_{TE}} \leq \eta_c$, with
$\eta_c= 1- \frac{T_L}{T_R}$ as the Carnot efficiency.
In the language of stochastic thermodynamics, 
corresponding stochastic variables can be defined, the results of measurements during the time interval $t$,
the fluctuating work $-w=-t\dot{w}$ and input heat flow $q=t\dot{q}$ \cite{espoRev}.
%
%
One can further define the stochastic efficiency for a single realization as $\eta_{TE}= -{w}/{q}$. 

In our formalism, we obtain the CGF for work and heat by
going back to the definition of the characteristic function, Eq. (\ref{eq:GF}).
Rather that using the counting fields $\lambda_p$ and  $\lambda_e$ for charge and energy, 
we make the following substitutions, to obtain cumulants for work and heat,
%
\bea
\lambda_e &\to& \lambda_q,  \nonumber \\
\lambda_p &\to& -\lambda_q  \mu_R - \lambda_w (\mu_L-\mu_R).
\label{eq:transform}
\eea
$\lambda_q$ and $\lambda_w$ are conjugate counting parameters 
for  $q=H_R-\mu_RN_R$ and $-w=(\mu_L-\mu_R)N_R$, respectively.
This transformation modifies the form of the fluctuation symmetry
\be
{\cal G}(\lambda_w,\lambda_q) = {\cal G}(-\lambda_w + i \beta_L, -\lambda_q+i(\beta_L -\beta_R )),
\ee
which immediately implies (using $\lambda_w=\lambda_q=0$)
that \cite{Sinitsyn, Lahiri, Campisi}
\be
\Big\langle \exp \Big[-\frac{w}{T_L} - \Big(\frac{1}{T_L}-\frac{1}{T_R}\Big) q \Big]\Big\rangle =1.
\label{work-flux}
\ee
By invoking the Jensen's inequality, Eq.~(\ref{work-flux}) immediately returns the bound
$-\langle w\rangle/\langle q \rangle\leq \eta_c$, confirming that our definitions for $q$ and $w$ are
consistent with classical thermodynamics.
In contrast, efficiency fluctuations are typically not bounded,
and can take arbitrary values because of the stochastic nature of small systems.
Therefore, in general, it is useful to investigate the probability distribution function 
to obtain the fluctuating work and heat within the interval $t$, thus he probability distribution function $P_t(\eta)$,
to observe the value $\eta$ within $t$. According
to the theory of large deviations, the probability function assumes an asymptotic long time form \cite{LDF},
%
\bea
P_t(\eta) \sim e^{-t\tilde J(\eta)}
\eea
with $\tilde J (\eta)$ identified as the ``large deviation function". 
We rescaled here and below the efficiency by the Carnot value, 
\bea
\eta\equiv \eta_{TE}/\eta_c.
\eea
The upper bound of the efficiency thus corresponds to the value $\eta=1$.

It can be shown \cite{Esp4,Esp1,Esp3} that the large deviation function for efficiency can be obtained from
${\cal G}(\lambda_w, \lambda_q)$ by setting $\lambda_q=\eta \,\eta_c \lambda_w$, and minimizing it with respect to
$\lambda_w$, 
\be
\tilde {J}(\eta) = - \min_{\lambda_w} {\cal G}(\lambda_w, \eta\, \eta_c\, \lambda_w).
\label{eq:tildeJ}
\ee
The CGFs for the AH and HO mode models are given in Eqs. (\ref{eq:CGF-AH}), (\ref{eq:CGF-H}), respectively.
To study efficiency fluctuations we use the transformation (\ref{eq:transform}), and receive the LDF from Eq. (\ref{eq:tildeJ}).
Note that we do not explicitly evaluate the probability distribution function $P_t(\eta)$.

It can be proved that $\tilde J(\eta)$ has a single minimum, coinciding with the macroscopic efficiency
of the engine, and a single maximum, corresponding to the least likely efficiency,
which equals to the Carnot efficiency, $\eta=1$ \cite{Esp2,Esp3,Esp4,Esp1}.

\begin{figure}
\includegraphics[width=8cm]{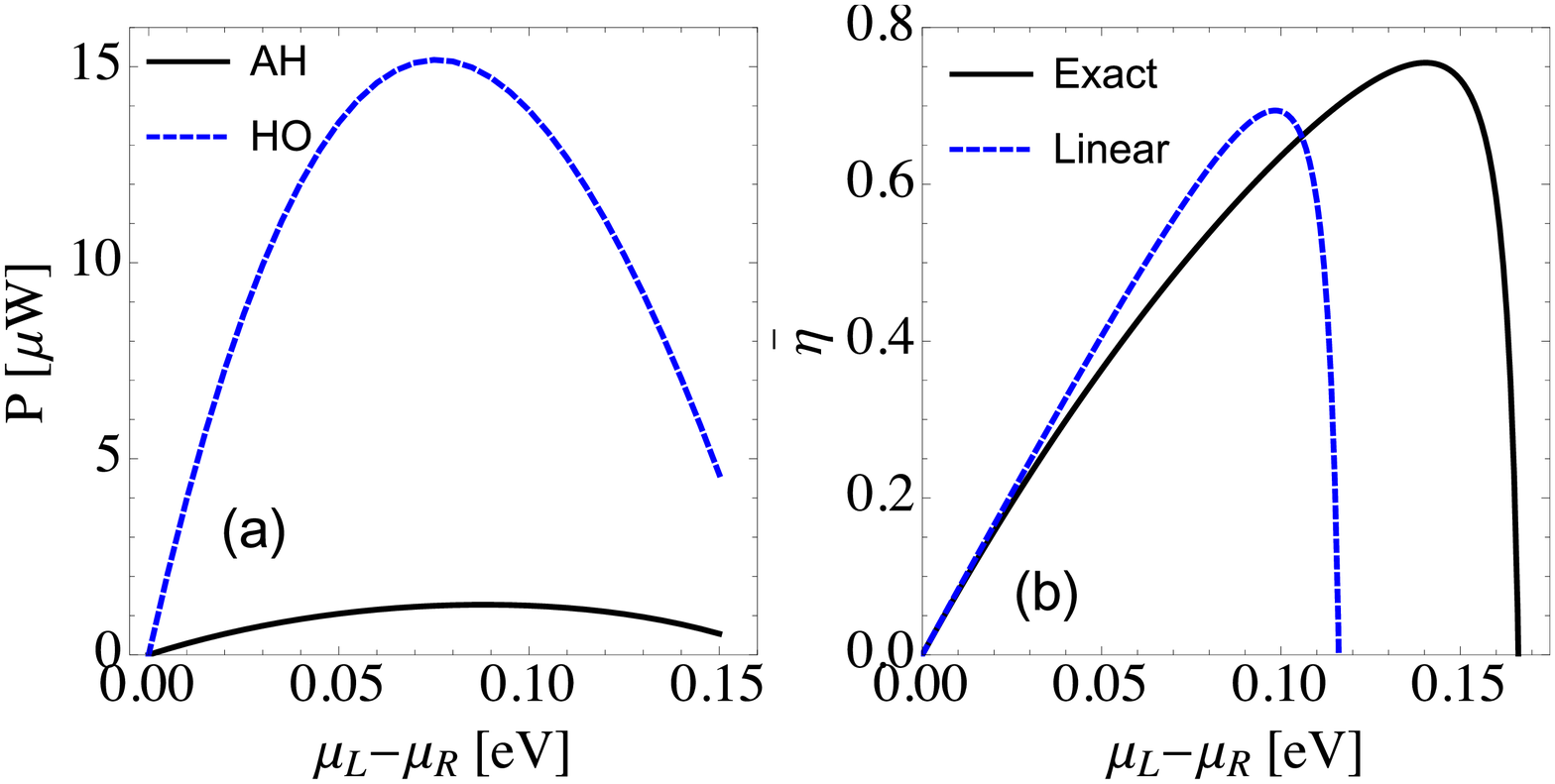}
\caption{(color online)
(a) Output power
$P= (\mu_L-\mu_R) \langle I_p\rangle$ and (b) macroscopic efficiency
$\bar \eta=\bar \eta_{TE}/\eta_c$, comparing exact results to the
value (\ref{eq:etaG}).
Parameters are $\epsilon_a=\epsilon_d=0.2$ eV,
$\omega_0=0.01$ eV, $g=0.1$ eV, $\Gamma=0.1$ eV, $T_L=300$ K, $T_R=800$ K, $\Gamma_{\rm ph}=0$.
For linear response calculations we define the average temperature as $T_a= (T_L+T_R)/2$,
the temperature difference $\Delta T= T_R-T_L$, and similarly for the chemical potential,
$\mu_a=(\mu_L+\mu_R)/2$ and $\Delta {\mu}=\mu_R-\mu_L$.}
\label{eff-power}
\end{figure}


\begin{figure*}
\includegraphics[width=13cm]{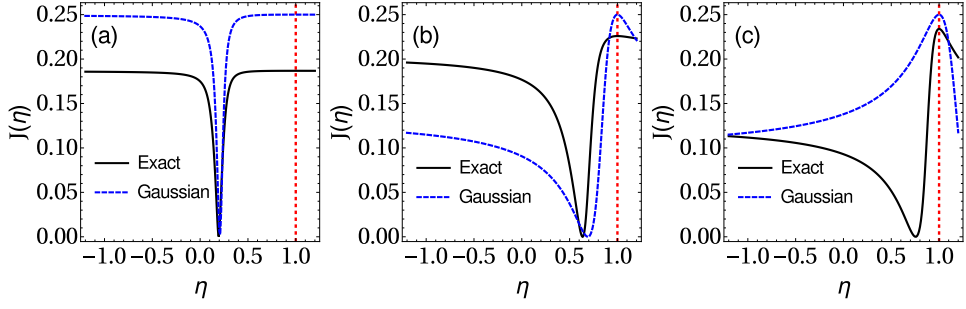}

\caption{(color online)
Efficiency LDF for the DA-AH model, showing the exact result
$J_{AH}(\eta)$ and the Gaussian limit $J_G^{AH}(\eta)$ at different biases.
Parameters are $\epsilon_a=\epsilon_d=0.2$ eV, $\omega_0=0.01$ eV, 
$g=0.1$ eV, $\Gamma=0.05$ eV, $\Gamma_{\rm ph}=0$, $T_L=300$ K, $T_R=800$ K.
(a) $\Delta \mu=0.025$ eV, (b) $\Delta \mu=0.1$ eV, (c) $\Delta \mu=0.14$ eV.
The vertical dashed line identifies the scaled Carnot efficiency $\eta_c$=1.}  
\label{J-AH}
\end{figure*}

\begin{figure*}
\includegraphics[width=13cm]{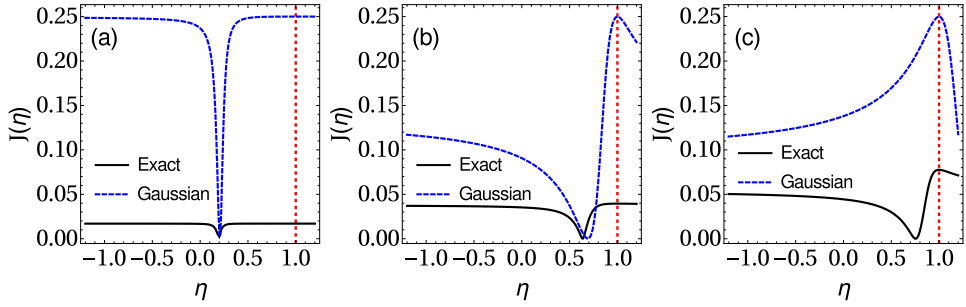}
\caption{(color online) Efficiency LDF $J_{HO}(\eta)$ and $J_G^{HO}(\eta)$ for HO model for different bias voltage.
Parameters are same as in Fig.~(\ref{J-AH})
with (a) $\Delta \mu=0.025$ eV, (b) $\Delta \mu=0.1$ eV, (c) $\Delta \mu=0.14$ eV.
The vertical dashed line identifies the scaled Carnot efficiency $\eta_c$=1.
}
\label{J-HO}
\end{figure*}

\begin{figure*}
\includegraphics[width=13cm]{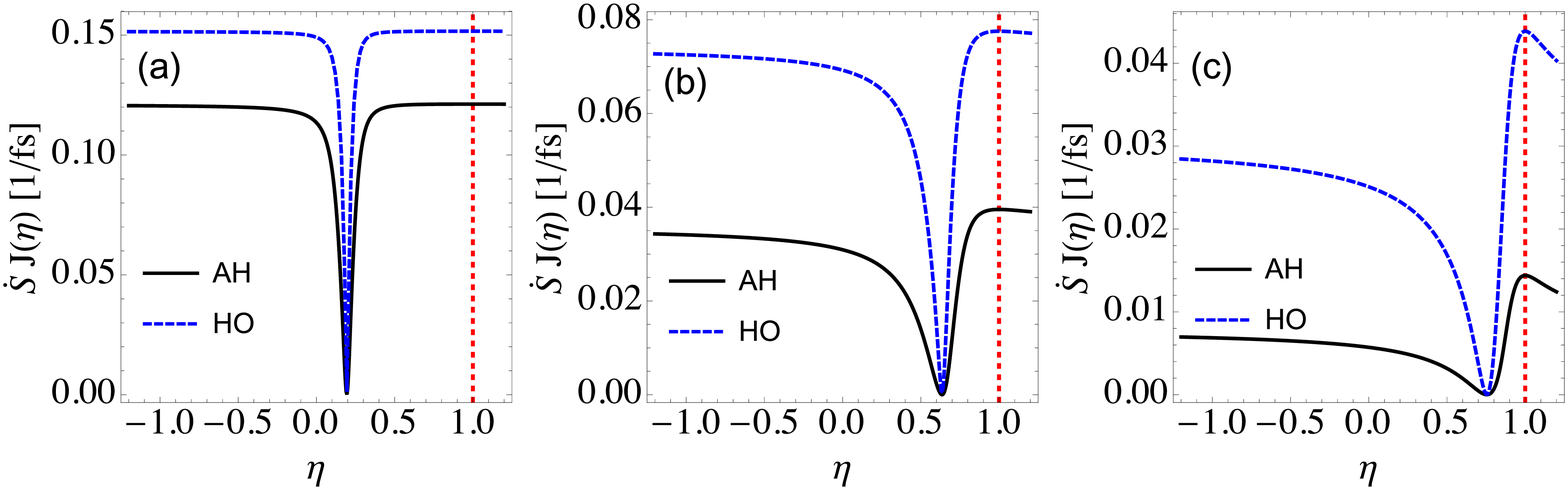}
\caption{(color online)
The exact LDF $\dot{S} J(\eta)$ (unnormalized)
for the DA-AH and the DA-HO models at different biases
(a) $\Delta \mu=0.025$ eV, (b) $\Delta \mu=0.1$ eV, (c) $\Delta \mu=0.14$ eV.
Parameters are same as in Fig.~(\ref{J-AH}).
The vertical dashed line identifies the scaled Carnot efficiency $\eta_c$=1.}
\label{J-eta-comp}
\end{figure*}



\subsubsection{Gaussian limit: Linear response theory}
\label{etaLR}

In the linear response limit, i.e. close to equilibrium, the stochastic work and heat are assumed
to be Gaussian variables.
It is possible then to derive an explicit expression for the large deviation function,
expressed in terms of the Onsager's response coefficients and the thermodynamic affinities \cite{Hua-Bijay, Esp1,Esp2,Esp3}. The scaled-dimensionless
LDF is defined as 
\bea
J(\eta)= \tilde J(\eta)/\dot{S},
\eea
with the entropy production rate $\dot S$. 
In the present Gaussian ($G$) limit it is given by \cite{Hua-Bijay, Esp1,Esp2}
\be
J_{G}(\eta)= \frac{1}{4} \frac{(\eta+ \alpha^2 + \alpha d + \alpha d \eta)^2}{(1+ \alpha^2 + 2 \alpha d) (\eta^2+ \alpha^2 + 2 \alpha d \eta)},
\label{eq:Gaussian-J}
\ee
%
with the dimensionless parameters
\be
d \equiv \frac{L_{pq}}{\sqrt{L_{pp} L_{qq}}}, \quad \alpha\equiv \frac{A_p \sqrt{L_{pp}}}{A_q \sqrt{L_{qq}}}.
\ee
Here $d$ describes the degree of coupling in the system, $\alpha$  is the affinity parameter.
Note that in the Gaussian limit $J_{G}(\eta)$ is bounded,
$0 \leq J_{G}(\eta) \leq \frac{1}{4}$. The minimum value $J_{G}(\bar{\eta}_G) =0$ is obtained at
the average (macroscopic) efficiency 
\bea
\bar{\eta}_G= -\frac{\alpha (\alpha + d)}{(1+ \alpha d)}.
\label{eq:etaG}
\eea
The maximum value $J_{G}(\eta_c=1)=1/4$
shows up precisely at the Carnot efficiency, $\eta_c=1$.

To simulate (\ref{eq:Gaussian-J})
we get hold of $d$ and $\alpha$ by extracting numerically the coefficients of the linear response-average 
charge $\langle I_p \rangle $ and heat currents $\langle I_q\rangle$, 
\be
\langle I_p \rangle = L_{pp} A_p + L_{pq} A_q, \quad  \langle I_q \rangle = L_{qp} A_p+ L_{qq} A_q.
\ee
Time-reversal symmetry guarantees that $L_{pq}=L_{qp}$.
The affinities responsible for the particle and heat fluxes
are $A_p=\beta_L (\mu_R-\mu_L)$ and $A_q=\beta_L-\beta_R$, respectively.
The average entropy production rate, valid in general non-equilibrium situations, is
$\dot{S}= \langle I_p \rangle  A_p + \langle I_q \rangle A_q$ \cite{EPR}.
%
Note that we plotted the dimensionless LDF in Figs. \ref{J-AH} and \ref{J-HO}, when presenting both exact and linear-response results.
Only in Fig. \ref{J-eta-comp} we retract to the unscaled function, when comparing different models.
%

\subsubsection{Numerical Results: efficiency statistics}
\label{etaR}

We investigate numerically the  thermoelectric efficiency and its statistics in the DA model,
considering the effect of mode harmonicity and beyond linear response situations.
We begin with macroscopic-averaged properties.
Inspecting  Eq. (\ref{eq:AH/HO-current}), we note that
the numerator in the expressions for the average charge and energy currents are identical in the DA-AH and the DA-HO models.
Since the macroscopic efficiency is proportional to the ratio of these two currents, we immediately conclude
that regardless of whether the mode is harmonic/two-state system,
the same macroscopic efficiency is to be reached, at an arbitrary non-equilibrium condition.
However, the output power takes different values in the two models.

In Fig. \ref{eff-power}(a)
we display the generated power $P = (\mu_L-\mu_R) \langle I_p \rangle $ as a function of  bias voltage
$\mu_L-\mu_R$ for both DA-AH and DA-HO models.
We find that when the mode is harmonic
the output power can largely exceed values reached in a junction with an AH mode due to
the availability of many additional channels.
In Fig. \ref{eff-power}(b) we examine the efficiency far from equilibrium, and compare the exact value to the
linear response limit.
For the given parameters, linear response theory agrees with the exact efficiency calculation
as long as $\Delta \mu \leq 0.03$ eV. Interestingly,
we find that the device can be made more efficient in the nonlinear regime,
 $(\bar \eta\approx 0.75$ at  $\mu_L-\mu_R \approx 0.14$ eV),
in contrast to the linear response limit (\ref{eq:etaG}), which is obtained by linearizing the currents around equilibrium,
 to extract the Onsager coefficients.
We further recall the scaling $P\propto g^2$, and that $\bar\eta$ itself does not depend on $g$.

We now turn our attention to the efficiency statistics.
In Fig. \ref{J-AH} we display the scaled LDF $J_{AH}(\eta)$ for the model with an AH mode,
(normalized by the entropy production rate $\dot{S}$).
It is obtained from Eq. (\ref{eq:tildeJ}) by minimizing the analytical form for the CGF
(\ref{eq:CGF-AH}) with respect to $\lambda_{w}$.
We further compare the exact LDF with the Gaussian limit, $J_G^{AH}(\eta)$, which is obtained
from Eq. (\ref{eq:Gaussian-J})
by linearizing (numerically) the currents, to obtain the parameters $d$ and $\alpha$, different in general for the DA-HO and DA-AH models.
$J(\eta)$ does not depend on $g$ given the normalization with the entropy production rate.

We examine the efficiency statistics in the three panels of Fig. \ref{J-AH}
at different-representative values for the applied voltage: (a) linear response limit $\Delta \mu=0.025$ eV,
(b) beyond linear response, $\Delta \mu=0.1$ eV, and (c) around the maximal value for efficiency, $\Delta \mu=0.14$ eV.
We find that the minimum value of $J_{AH}(\eta)$ corresponds to the macroscopic efficiency,
and that the Carnot efficiency $\eta=1$ is the least-likely efficiency. We also confirm that
$J_G^{AH}(\eta)$ is always bounded between 0 and 1/4, with the upper bound reached precisely
at the Carnot efficiency $\eta=1$.
In contrast, we find (numerically) that at $\eta=1$ the exact LDF satisfies
$J(1) \leq 1/4$; equality is reached
only in the linear response limit.
When increasing the bias voltage the magnitude of efficiency fluctuations grows,
and, as expected, the Gaussian approximation $J_G(\eta)$ becomes increasingly unreliable.


In Fig. \ref{J-HO} we display the LDF for the DA-HO model.
The observed trends are similar to those of Fig. \ref{J-AH}.
However, because of the large entropy production rate taking place in the DA-HO junction,
the normalized $J(\eta)$ receives rather low values.

We further compare the two models for the vibrational mode
and plot the unnormalized LDF $\dot{S} J(\eta)$ in Fig.~(\ref{J-eta-comp}).
As stated before, the macroscopic efficiency coincides in these models.
However, quite interestingly, the overall statistics differs when increasing bias.
In the linear regime, the DA-HO model performs as an effective two-state system at low temperatures
 and deviations only show up at the tail of the distribution.
At high bias, the  DA-AH model suffers more significant efficiency fluctuations relative to the DA-HO model.

We conclude this section emphasizing central observations: 
The statistics of efficiency can reveal information on the mode harmonicity,
the Gaussian-linear response limit becomes highly unreliable at large bias, as expected.


\section{Conclusions}
\label{Summ}

We provided a comprehensive analysis of
 vibrationally assisted charge and energy transport in a donor-acceptor type molecular junction.
Two limiting models were examined: (a) In the DA-AH junction the vibrational mode was highly anharmonic,
consisting of a two-level system. (b) The mode was taken as harmonic in the DA-HO model.
Key results are:


(i) 
Employing QME and NEGF approaches for the DA-AH and the DA-HO models, respectively,
we obtained analytical expressions for the steady state cumulant generating functions,
Eqs. (\ref{eq:CGF-AH}) and (\ref{eq:CGF-H}).
These results are valid to second-order in the electron-phonon strength, and correct to arbitrary order 
in the molecule-metal coupling.
The CGF furnishes analytical results for charge and energy currents in the system, and for
fluctuations of these quantities.

(ii) 
Our analysis establishes that one can reconcile
two different-eminent quantum transport techniques:  QME and NEGF.
By taking into account scattering processes to the same order in perturbation theory,
we showed that the QME (used here for treating the DA-AH junction)
and the NEGF approach (DA-HO model) yielded corresponding results.
Several works had compared transport predictions from these two approaches,
showing deviations given the different approximations involved \cite{Esp1,Peskin}. 
Our work here is unique in demonstrating
that one can reconcile results from these two techniques by carefully taking into account corresponding processes.

(iii) 
Expressions for the current, output power,
as well as the Fano factor (noise) were reached, shown to
be sensitive to the properties of the vibrational mode.
Specifically, we found that in our model the two-peak structure of the differential conductance
directly evinces on the mode frequency,
while the Fano factor definitely reveals information on the mode harmonicity/anharmonicity.
In contrast, the macroscopic (averaged) thermoelectric efficiency was proved to be
identical regardless of the mode harmonicity, though
fluctuations around the averaged value were distinct in the two cases.

(iv) 
We had employed the DA junction as a thermoelectric engine and studied its
 efficiency fluctuations based on the derived CGF for charge and energy transfer.
Previous works of efficiency fluctuations were limited to classical models \cite{Esp2,Esp3,Esp4},
or to the quantum - purely coherent (noninteracting) regime \cite{Esp1}.
Here, in contrast, we examined efficiency fluctuations far from equilibrium
in a quantum many-body model, using a rigorous approach.

In future studies we will demonstrate the correspondence
between the QME an NEGF in the DA-HO model, and examine energy harvesting in a double-dot cell
with three terminals, to examine the quantum photovoltaic effect.

\section*{Acknowledgments}
The work of DS and BKA was supported by an NSERC Discovery Grant, the Canada Research Chair program,
and the CQIQC at the University of Toronto.
JHJ acknowledges support from the faculty start-up funding of Soochow University.

\renewcommand{\theequation}{A\arabic{equation}}
\setcounter{equation}{0}  
\section*{Appendix A: Real time Green's functions}
\subsection{Free electron Green's function}

The free electronic Green's functions for the leads, with the contour times $\tau_1$ and $\tau_2$, are given as
\bea
g_k(\tau_1,\tau_2)= -i \,\langle T_c a_{k}(\tau_1) a_{k}^{\dagger}(\tau_2)\rangle, \quad k\in L,R
\eea
The projection of this contour ordered Green's function to real time generates four different components,
namely, lesser ($<$), greater $(>$), time-ordered ($t$) and anti-time ordered ($\bar{t}$) Green's functions.
The lesser and greater components are given as (e.g., for $l\in L$),
\bea
g^{<}_{l}(t_1-t_2) &=& i \, \langle a_{l}^{\dagger}(t_2) a_{l}(t_1)\rangle= i f_{L}(\epsilon_{l})
e^{i \epsilon_{l} (t_2-t_1)}, \nonumber \\
g^{>}_{l}(t_1-t_2) &=& - i\,\langle a_{l}(t_1) a_{l}^{\dagger}(t_2)\rangle=
- i [1-f_{L}(\epsilon_{l})] e^{i \epsilon_{l} (t_2-t_1)}.
\nonumber\\
\eea
In frequency domain we get
\bea
g^{<}_{l}(\omega) &=& 2 \pi i f_{L}(\epsilon_{l}) \delta(\omega-\epsilon_{l}),
\nonumber\\
g^{>}_{l}(\omega) &=& - 2 \pi i [1-f_{L}(\epsilon_{l})] \delta(\omega-\epsilon_{l}).
\eea
The following relations between different components of the Green's functions are valid in both time and frequency domain:
\bea
g_{l}^t&=& g_{l}^r + g_{l}^< = g_{l}^a + g_l^>,
\nonumber\\
g_{l}^{\bar t}&=& g_{l}^< - g_{l}^a = g_{l}^> - g_{l}^r,
\label{electron-relation}
\eea
where $g_{l}^t$ and $g_{l}^{\bar t}$ are time-ordered and anti time-ordered Green's functions.
The retarded Green's function is defined as
\bea
g_{l}^r(t_1,t_2)=-i\Theta(t_1-t_2)\langle\{ a_{l}(t_1), a^{\dagger}_{l}(t_2)\} \rangle,
\eea
and the advanced Green's function is $g_{l}^a(t_1,t_2)=[g_{l}^r(t_2,t_1)]^{*}$.
In a similar manner, counting field dependent Green's functions are defined on the contour as
\bea
\tilde{g}_{r}(\tau_1,\tau_2)= - i \,\langle T_c \tilde{a}_{r}(\tau_1) \tilde{a}_{r}^{\dagger}(\tau_2)\rangle,
\eea
where we employ the short notation 
$\tilde{a}_{r}(\tau)\equiv e^{ -i (\lambda_p(\tau) + \epsilon_{r} \lambda_e(\tau))} a_{r}(\tau)$.
In real time, we obtain the lesser and greater components,
\bea
\tilde{g}^{<}_{r}(t_1\!-\!t_2) &=& i \langle \tilde{a}_{r}^{\dagger}(t_2) \tilde{a}_{r}(t_1)\rangle
\nonumber\\
&=& i \, f_{R} (\epsilon_r) e^{i \epsilon_{r} (t_2-t_1)} e^{i (\lambda_p+ \epsilon_{r} \lambda_e)},
\nonumber \\
\tilde{g}^{>}_{r}(t_1\!-\!t_2) &=& - i  \langle \tilde{a}_{r}(t_1) \tilde{a}_{r}^{\dagger}(t_2)\rangle
\nonumber\\
&=& -i\, [1-f_{R}(\epsilon_r)] e^{i \epsilon_{r} (t_2-t_1)}\,e^{-i (\lambda_p+ \epsilon_{r} \lambda_e)}.
\nonumber\\
\label{counting-GF}
\eea
In frequency domain they are given by
\bea
\tilde{g}^{<}_{r}(\omega) &=& 2 \pi i f_{R}(\epsilon_{r}) \delta(\omega-\epsilon_{r}) e^{i (\lambda_p+ \epsilon_{r} \lambda_e)},\nonumber \\
\tilde{g}^{>}_{r}(\omega) &=& - 2 \pi i [1-f_{R}(\epsilon_{r})] \delta(\omega-\epsilon_{r}) e^{-i (\lambda_p+ \epsilon_{r} \lambda_e)}.
\label{counting-GF-frequency}
\eea
Note that relations such as in Eq.~(\ref{electron-relation}) do not hold for counting-field dependent Green's functions.

\subsection{Electron-hole Green's function}

In the main-text  we have defined  the electron-hole propagator in Keldysh space [Eq.~(\ref{eh-prop})].
In real time we receive the four different components as
\bea
F^t(t_1,t_2) &=& -i g^2 \sum_{l,r} |\gamma_l|^2 |\gamma_r|^2
\nonumber\\
&\times&
\big[g_{l}^t (t_1\!-\!t_2)\, g_{r}^t (t_2\!-\!t_1)+ g_{l}^t (t_2\!-\!t_1)\, g_{r}^t (t_1\!-\!t_2)\big],
\nonumber \\
F^{\bar t}(t_1,t_2) &=& -i g^2 \sum_{l,r} |\gamma_l|^2 |\gamma_r|^2
\nonumber\\
&\times&
\big[g_{l}^{\bar{t}} (t_1\!-\!t_2)\, g_{r}^{\bar{t}} (t_2\!-\!t_1)+ g_{l}^{\bar{t}} (t_2\!-\!t_1)\, g_{r}^{\bar{t}} (t_1\!-\!t_2)\big],
\nonumber \\
\tilde{F}^<(t_1,t_2) &=& -i g^2 \sum_{l,r} |\gamma_l|^2 |\gamma_r|^2
\nonumber \\
&\times&
\big[g_{l}^{<} (t_1\!-\!t_2)\, \tilde{g}_{r}^{>} (t_2\!-\!t_1)+ g_{l}^{>} (t_2\!-\!t_1)\, \tilde{g}_{r}^{<} (t_1\!-\!t_2)\big],
\nonumber \\
\tilde{F}^>(t_1,t_2) &=& -i g^2 \sum_{l,r} |\gamma_l|^2 |\gamma_r|^2
 \nonumber\\
 &\times&\big[g_{l}^{>} (t_1\!-\!t_2)\, \tilde{g}_{r}^{<} (t_2\!-\!t_1)+ g_{l}^{<} (t_2\!-\!t_1)\, \tilde{g}_{r}^{>} (t_1\!-\!t_2)\big].
\nonumber\\
\eea
In frequency domain assuming time-translational invariance for the propagator in the steady state limit,
these components can be written as
\begin{widetext}
\bea
F^t (\omega) &=&- i g^2 \sum_{l,r} |\gamma_l|^2 |\gamma_r|^2 \int_{-\infty}^{\infty} \frac{d\omega'}{2\pi} \big[g_{l}^t (\omega_+)\, g_{r}^t (\omega_-)+ g_{l}^t (\omega_-)\, g_{r}^t (\omega_+)\big],  \nonumber \\
F^{\bar t} (\omega) &=& -i g^2 \sum_{l,r} |\gamma_l|^2 |\gamma_r|^2 \int_{-\infty}^{\infty} \frac{d\omega'}{2\pi} \big[g_{l}^{\bar t} (\omega_+)\, g_{r}^{\bar t} (\omega_-)+ g_{l}^{\bar t} (\omega_-)\, g_{r}^{\bar t} (\omega_+)\big], \nonumber \\
\tilde{F}^{<} (\omega) &=& -i g^2 \sum_{l,r} |\gamma_l|^2 |\gamma_r|^2 \int_{-\infty}^{\infty} \frac{d\omega'}{2 \pi} \big[g_{l}^{<} (\omega_+)\, \tilde{g}_{r}^{>} (\omega_-)+ g_{l}^{>} (\omega_-)\, \tilde{g}_{r}^{<} (\omega_+)\big],  \nonumber \\
\tilde{F}^{>} (\omega) &=& -i g^2 \sum_{l,r} |\gamma_l|^2 |\gamma_r|^2 \int_{-\infty}^{\infty} \frac{d\omega'}{2\pi} \big[g_{l}^{>} (\omega_+)\, \tilde{g}_{r}^{<} (\omega_-)+ g_{l}^{<} (\omega_-)\, \tilde{g}_{r}^{>} (\omega_+)\big],
\eea
\end{widetext}
where $\omega_{\pm}=\omega' \pm \frac{\omega}{2}$.
Using the relations between the Green's functions [Eq.~(\ref{electron-relation})] and the expressions for the free Green's
functions [Eq.~(\ref{counting-GF-frequency})] we obtain the lesser and greater components for the propagator,
\begin{widetext}
\bea
\tilde{F}^{<} (\omega) &=& -i 2 \pi g^2 \Big[ \sum_{l,r} |\gamma_l|^2 |\gamma_r|^2 f_L(\epsilon_l) (1-f_R(\epsilon_r)) e^{-i(\lambda_p+\epsilon_r \lambda_e)} \delta(\epsilon_l-\epsilon_r-\omega) \nonumber \\
&+&  \sum_{l,r} |\gamma_l|^2 |\gamma_r|^2 f_R(\epsilon_r) (1-f_L(\epsilon_l)) e^{i(\lambda_p+\epsilon_r\lambda_e)} \delta(\epsilon_l-\epsilon_r+\omega)\Big],
\eea
\end{widetext}
and $\tilde{F}^>(\omega)= \tilde{F}^<(-\omega)$.
Note that at the phonon frequency $\omega_0$ the lesser and greater components
of $\tilde{F}$ are related to the excitation and relaxation rates,
defined in the QME approach, as $\tilde{F}^<(\omega_0) =-i \, k_{u}^{\lambda}$ and
$\tilde{F}^>(\omega_0) =-i \, k_{d}^{\lambda}$.

The sum of $t$ and $\bar{t}$ components can be simplified following the relations
(\ref{electron-relation}), and using the identity $\int_{\infty}^{\infty} d\omega' g^{r,a}(\omega_+) g^{r,a}(\omega_{-})=0$,
to reach
\begin{widetext}
\bea
&&F^t (\omega) + F^{\bar{t}}(\omega)= -i g^2 \sum_{l,r} |\gamma_l|^2 |\gamma_r|^2
\int_{-\infty}^{\infty} \frac{d\omega'}{2 \pi} \big[g_{l}^{<} (\omega_+)\, {g}_{r}^{>} (\omega_-)+ g_{l}^{>} (\omega_-)\, {g}_{r}^{<} (\omega_+) + g_{l}^{>} (\omega_+)\, {g}_{r}^{<} (\omega_-)+ g_{l}^{<} (\omega_-)\, {g}_{r}^{>} (\omega_+)\big]
\nonumber \\
&&= F^>(\omega)+F^<(\omega)
\eea
\end{widetext}
%
We further obtain
\be
F^t(\omega)-F^{\bar{t}}(\omega)= 2 \,{\rm Re} \big[F^r(\omega)\big],
\ee
where
\bea
F^r(\omega) &=&
- ig^2 \sum_{l,r}|\gamma_l|^2|\gamma_r|^2 \int d\omega' \Big \{ {\rm Re}[g_l^r(\omega_+)] g_r^{<}(\omega_{-})
\nonumber\\
&+& {\rm Re}[g_r^r(\omega_-)] g_l^{<}(\omega_{+}) \Big\} + {l \leftrightarrow r}
\nonumber\\
\eea

\renewcommand{\theequation}{B\arabic{equation}}
\setcounter{equation}{0}  
\section*{Appendix B: Generating function in the presence of a phonon bath}

The CGFs derived for the DA-AH and DA-HO models can be extended to describe the case with a vibrational mode linearly coupled to
a dissipative phonon bath. Assuming this coupling to be weak, it can be shown that
the formal expressions for the CGFs, Eqs. (\ref{eq:CGF-AH}) and (\ref{eq:CGF-H}), remain the same
except that the excitation and relaxation rates are modified, given now by the sum of electronic-baths and 
phononic-bath induced contributions,
%
%
\bea
k_{u}^{\lambda}  &=& k_{u}^{\lambda, \rm el}+  k_{u}^{\rm ph}, \nonumber \\
k_{d}^{\lambda}  &=& k_{d}^{\lambda,\rm el} +  k_{d}^{\rm ph},
\eea
with
\bea
k_{u}^{\rm ph}  &=& \Gamma_{\rm ph}(\omega_0) n_{\rm ph}(\omega_0),  \nonumber \\
k_{d}^{\rm ph} &=&  \Gamma_{\rm ph}(\omega_0) [1+ n_{\rm ph}(\omega_0)].
\eea
$k_{u,d}^{\lambda,\rm el}$ are the rates defined in the main text, induced by the metal leads.
Here, $\Gamma_{\rm ph}(\omega_0)$ is the coupling energy of the particular mode $\omega_0$ 
to the phonon bath and $n_{ \rm ph} (\omega)= [\exp(\beta_{ \rm ph} \omega)-1]^{-1}$
is the Bose-Einstein distribution function at temperature $1/\beta_{\rm ph}$.
Note that in the presence of the additional phonon bath the fluctuation symmetry as written in
Eq.~(\ref{eq:fluc-sym}) is not satisfied.
To restore the symmetry, one should 'count' as well the energy dissipated into the phonon bath.

For completeness, we include the expressions for the charge current and its noise in the 
dissipative harmonic mode model,
generalizing Eqs. (\ref{eq:AH/HO-current}) and (\ref{eq:AH/HO-noise}),
\bea
\langle I_p ^{HO}\rangle =\frac{(k_u^{\rm el})' k_d  + (k_d^{\rm el})'k_u}{k_d-k_u},
\eea
with the short notation $(k_u^{\rm el})'\equiv [k_u^{\rm el}]^{R \to L} - [k_u^{\rm el}]^{L \to R}$ and
$(k_d^{\rm el})'\equiv [k_d^{\rm el}]^{R \to L} - [k_d^{\rm el}]^{L \to R}$.
The charge current noise is
\bea
\langle S_p^{HO}\rangle  =
2 \frac{\langle I_p^{HO}\rangle^2}{k_d-k_u}
+\frac{k_u^{\rm el} k_d +k_d^{\rm el} k_u +2 (k_u^{\rm el})'(k_d^{\rm el})'}{k_d-k_u}.
\nonumber\\
\eea
The second term clearly indicates that the noise includes terms mixing the effects of the three reservoirs.
These expressions were used to simulate Figures \ref{current-H} and \ref{fano-H}, taking $\Gamma_{\rm ph}$
as a constant.

\begin{thebibliography}{99}
\bibitem{commM}
Following the two-time measurement procedure,
the average in Eq. (\ref{eq:GF}) [and similarly Eq. (\ref{eq:ZGF})] 
should be defined with respect to a modified density operator
$\rho'_T(0)=\sum_{a} \Pi_a \rho_T(0) \Pi_a$, where $\Pi_a$ are projection operators,
prepared from the eigenvectors of
the measuring operators $N_R, H_R$ \cite{Huanan}. 
However, for the choice of product initial state 
$\rho_T(0)= \rho_{vib}(0) \otimes \rho_L\otimes \rho_R$, as employed here,
the projection operator $\Pi_a$ commutes with $\rho_T(0)$, therefore
$\rho'_T(0)=\rho_T(0)$, leading to Eqs. (\ref{eq:GF}) and 
(\ref{eq:ZGF}).
\bibitem{commentSF}
In Ref. \cite{SegalSF} we derived the cumulant generating function of the DA-AH
junction by energy resolving the quantum master equation for the population of the vibrational mode.
Here we perfom this derivation rigorously by using a counting statistics approach
from the principles of the projection operator technique.
We also confirmed in Ref. \cite{SegalSF} that the CGF of the DA-HO junction obeyed
the fluctuation symmetry, but did not
achieve the CGF itself. It is derived in this work for the first time using the NEGF method.
\bibitem{comment2}
Figures \ref{current-H} and \ref{fano-H} were generated by emplying an additional phonon bath.
The current is scaled down by a factor $c$
if we scale parameters as $g/\sqrt{c}$ and $\Gamma_{\rm ph}/c$.
\bibitem{EPR}
The entropy production rate is identified  in steady state
by calculating the balance of entropy exchange at the terminals,
$\dot S=\beta_{L} \left(\langle I_e\rangle  -\mu_L\langle I_p\rangle \right)
- \beta_{R} \left(\langle I_e\rangle  -\mu_R\langle I_p\rangle\right)$.
Rearranging this expression, we can write it
in the standard form $\dot S=A_p \langle I_p\rangle +A_q \langle I_q\rangle$.
\end{thebibliography}
\end{document}